\documentclass{article}
\usepackage{graphicx}
\title{ROTATION RATES OF THE CORONAL HOLES AND THEIR PROBABLE ANCHORING DEPTHS}
\author{K. M. Hiremath and Hegde, M.\\ 
Indian Institute of Astrophyscis, Bengaluru-560034}

\begin{document}
%\author{ Gurman, J. B}
%\affil{NASA Goddard Space Flight Center, Greenbelt, MD, USA}
%\author{K. M. Hiremath and Hegde, M.}
%\address{Indian Institute of Astrophysics, Bangalore, India}
\maketitle
\begin{abstract}
{For the years 2001-2008, we use full-disk, SOHO/EIT  195 $\AA$  
calibrated  images to determine latitudinal and day to day 
variations of the rotation rates of coronal holes. We estimate the
weighted average of heliographic coordinates such as latitude 
and longitude from the central meridian on the observed solar disk. For 
different  latitude zones between $40^{o}$ north - $40^{o}$ south, we compute
rotation rates, and find that, 
irrespective of their area, number of days observed on the solar disk and latitudes,
coronal holes rotate rigidly. Combined for all the latitude zones, 
we also find that coronal holes rotate rigidly during their evolution history. 
In addition, for all latitude zones, coronal holes
follow a rigid body rotation law during their first appearance.   
Interestingly, average first rotation rate ($\sim 438 \, nHz$) of the coronal holes,
computed from their first appearance on the solar disk,
match with rotation rate of the solar interior only below the tachocline.
}

\end{abstract}
%\maketitle
%\keywords{sun:rotation, sun:coronal holes, sun:rotation rates, sun:heliosesimically inferred
%rotation rate}

\section{INTRODUCTION}
Solar coronal holes (CH) are large regions in the solar corona
with low density plasma (Krieger et al. 1973;
Neupert \& Pizzo, 1974; Nolte et al. 1976; Zirker 1977; Cranmer 2009 and references therein; Wang 2009) and
unipolar magnetic field structures (Harvey \& Sheeley 1979; Harvey et al. 1982), 
distinguished as dark features in EUV and X-ray wavelength regimes. During the solar maximum,
CH are distributed at all latitudes, while at solar minimum, CH
mainly occur near the polar regions (Madjarska \& Wiegelmann 2009). In addition
to sunspot activity and magnetic activity
phenomena that strongly influence the Earth's
climate (Hiremath 2009 and references there in),
there is increasing evidence that, on short time scales,
occurrences of solar coronal holes trigger responses in the Earth's upper atmosphere and magnetosphere
(Soon et al. 2000; Lei et al. 2008;
Shugai et al. 2009; Sojka et al. 2009;
Choi et al. 2009;  Ram et al. 2010; Krista 2011; Verbanac et al. 2011).

Physics of solar cycle and activity phenomena
is not well understood (Hiremath 2010 and references therein).
In order to understand the solar cycle and activity
phenomena, an understanding of rotational structure of the solar interior and the
surface are necessary. On the other hand, rotation rate of the interior
and the surface are coupled with the rotation rate of the
solar atmosphere, especially the corona. Although
there is a general consensus regarding the interior rotation as inferred from the helioseismology (Dalsgaard \& Schou 1988; Thompson et al. 1996; Antia et al. 1998; Thompson et al. 2003 and references therein;
Howe 2009; Antia \& Basu 2010),
surface rotation rates as derived from sunspots (Newton \&  Nunn 1951; Howard et al. 1984;
Balthasar et al. 1986; Shivaraman et al. 1993; Javaraiah 2003),
Doppler velocity (Howard \& Harvey 1970; Ulrich et al. 1988; Snodgrass \& Ulrich 1990)
and magnetic activity features (Wilcox \& Howard 1970; Snodgrass 1983; Komm et al. 1993),
there is no such consensus (see also Li {\em et al.} 2012) on the magnitude and form of rotation law
for features in the corona.

For example, by using coronal holes as tracers (Wagner 1975; Wagner 1976; Timothy \& Krieger 1975; Bohlin 1977) and large
scale coronal structures (Hansen et al. 1969; Parker et al. 1982; Fisher \& Sime 1984; Hoeksema 1984; 
Wang et al. 1988; Weber et al. 1999; Weber \& Sturrock 2002),
previous studies show that corona rotates rigidly 
while other studies (Shelke \& Pande 1985; Obridko \& Shelting 1989; Navarro-Peralta
\& Sanchez-Ibarra 1994; Insley et al. 1995)
indicate differential rotation. In addition to using coronal holes as tracers, X-ray bright points (Chandra et al. 2010; Kariyappa 2008; Hara 2009), coronal bright points (Karachik et al. 2006; Braj\v{s}a et al. 2004; W\"{o}hl et al. 2010), and
SOHO/LASCO  images have been used for the computation of rotation rates and yield a differentially rotating corona. 
Recent studies using radio images at 17 GHz (Chandra et al. 2009) and synoptic observations of the O VI 1032 $\AA$ spectral line from the SOHO/UVCS telescope (Mancuso \& Giordano 2011), however, suggest that the corona rotates rigidly. As part of an ISRO (Indian
Space Research Organization) funded project, the present
study utilizes SOHO/EIT 195 $\AA$
calibrated  images for understanding the following four objectives :
(i) to check for latitudinal dependency of rotation rates of the coronal holes,
(ii) to study  rotation rates of CH during their first appearance on the observed disk,
(iii) irrespective of their latitude, to study day to day variation of rotation
rates of coronal holes and, (iv) to estimate probable anchoring
depths of coronal holes.
In section 2, we present the data used and method of analysis, and the results
of that analysis in section 3. In section 4, we present the
discussion on cause for rigid body rotation rate of
the coronal holes and estimate their probable anchoring depths with our conclusions.

\section{DATA AND ANALYSIS}
For the period  2001 to 2008, we use full-disk 
SOHO (Solar and Heliospheric Observatory)/EIT
images (Delaboudini\'{e}re {\em et al}. 1995) that have a resolution of 2.6 arc sec. per pixel in a bandpass around 195 $\AA$ to 
detect coronal holes. The period studied includes both intense activity near solar maximum and
the descent of solar activity parameters such as 10.7 cm flux to values of
$\sim$ half of their values around that maximum.
The obtained images are in FITS format and individual pixels are in units of
data number (DN). DN is defined to be
output of the instrument electronics  which
corresponds to the incident photon signal  converted into charge within
each CCD pixel (Madjarska \& Wiegelmann 2009).

 We consider coronal holes that appear and disappear between $40^{o}$ north - $40^{o}$ south latitude of the visible solar hemisphere. Using the SolarSoft eit\_prep routine (Freeland \& Handy 1998), 
we background subtracted, flat-fielded, degridded and normalized the images. 
As this calibration involves exposure normalization of the images, now onwards unit of DN is DN/sec.
We used the occurrence dates and position of CH from the ``{\em spaceweather.com''} website. 
As the  ``{\em spaceweather.com''} is not designed for scientific
use, we use readily available occurrence dates of CH only. By using approximate
position (heliographic coordinates) of CH from this website, we separate a 
region from the SOHO/EIT images for further analysis and extraction of relevant
physical parameters as described below.
CH is also confirmed if it has a bimodal distribution in the
intensity histogram.
  
In order to extract physical parameters of CH from the EIT images, we 
use FV interactive FITS file editor (http://heasarc.gsfc.nasa.gov/docs/software/ftools/fv/).
Depending upon shape of the CH, from the FV editor, a circle or an ellipse is
drawn covering the whole region of CH and, average DN (intensity) (that is
set as a threshold for detecting the boundary) of CH is computed for 
detecting the boundary (private communications with Prof. Aschwanden).
Similar to Karachik \& Pevtsov (2011), for some of the coronal holes, threshold
is modified to match the visually estimated boundary. 
This method yields results consistent with the
previous intensity histogram methods (Krista \& Gallagher 2009; Krista 2011; de Toma 2011 and
references there in). After determining the boundary of CH,
we employed SolarSoft coordinate routines to compute the
central meridian distance ($l_{i}$) (heliographic longitude
from the central meridian) and
latitude ($\theta_{i}$) of individual pixels within the CH.

Fig 1(a) shows a full disk, solar image with a typical CH 
close to the center and in the north-east quadrant, while Fig 1 (b) 
represents the same CH with its threshold DN contour map.
In Fig 2(a), DN histogram of the CH is presented. The bimodal distribution 
in the histogram confirms the DN values in the CH region (Krista \& Gallagher 2009; Krista 2011).
We summed the total number of pixels and total DN within the CH boundary, which in turn allowed us to compute
average heliographic coordinates such as latitude ($\theta$) and central meridian distance (L) of CH as follows:

\begin{equation}
       \theta = \frac{\displaystyle\sum\limits_{i=1}^n\theta_{i}*DN_{i}}{\displaystyle\sum\limits_{i=1}^n DN_{i}} \, \, \, \, \, \, \, \, \,
L=\frac{\displaystyle\sum\limits_{i=1}^n l_{i}*DN_{i}}{\displaystyle\sum\limits_{i=1}^n DN_{i}} \, ,
\end{equation}

\noindent where $\theta_{i}$, $l_{i}$, and $DN_{i}$ (for $i=1,n$, $n$
is number of pixels) are the latitude, the central meridian distance,
and DN values of individual pixels. This method of finding the average
heliographic coordinates of CH is equivalent
to a method in physics of finding the center of mass of an arbitrary
geometrical shape.

\begin{table*}
%\centering
%\begin{minipage}{220mm}
%\begin{center}
\caption{Computation of heliographic coordinates with different weights in equation 1.}
\hskip 8ex
 \begin{tabular}{@{}|lll| ll| ll| @{}}
 \hline
\multicolumn{3}{|c|}{CH1} & \multicolumn{2}{|c|}{CH2} & \multicolumn{2}{|c|}{CH3}\\ \hline
%                    CH1 &     CH2   & CH3                                 

Weights     & Longitude &   Latitude  & Longitude & Latitude   & Longitude & Latitude \\ 
           &  (Degree)  &  (Degree)   & (Degree)  & (Degree) & (Degree) & (Degree)        \\

 \hline
DN         & -8.374 & 12.604 &8.352  & -11.727 &  4.011 & -3.898 \\ \hline
1/DN & -8.445 & 12.836 &8.637  & -12.682 & 3.818 & -3.996 \\ \hline
Average    & -8.400 & 12.715 &8.489  & -12.206 & 3.917 & -3.945 \\ \hline

\end{tabular}
%\end{minipage}
%\end{center}
\end{table*}

As the average heliographic coordinates of CH are weighted by the intensity
(DN counts)  of the relevant pixel, thus one can argue that more weight 
is given to brighter pixels. However, this argument can not be valid as  
the intensity is weighted in the denominator also (see above equation 1)
and, hence, whatever
higher weights given to the brighter pixels in the numerator are also equally
compensated by the higher weights in the denominator. We also checked
with another weighting that emphasizes areas darker than the image mean, 
$(i.e., (\sum\limits_{i=1}^n DN_{i}/N)-DN_{i}))$ and obtained the 
same results of average heliographic coordinates suggesting
that weighted average used in equation (1) is correct
and is not biased towards the brighter pixels. 

For computation of heliographic coordinates
of CH, we also used weights with inverse of DN (1/DN) and
without weights ({\em i.e.,} simple averages) in equation 1 and
the results for three typical CH are presented in Table 1. Negative sign for the longitude
indicates the CH that are on the eastern side of the
central meridian and negative sign for the latitudes indicates the
CH that are in the southern hemisphere. 
One can notice from this table that 
irrespective of weighted and non-weighted averaging, 
computed heliographic coordinates of CH are nearly same.

\begin{figure}
\begin{center}
     Fig 1(a) \hskip 40ex  Fig 1(b)
    \begin{tabular}{cc}
      {\includegraphics[width=18pc,height=18pc]{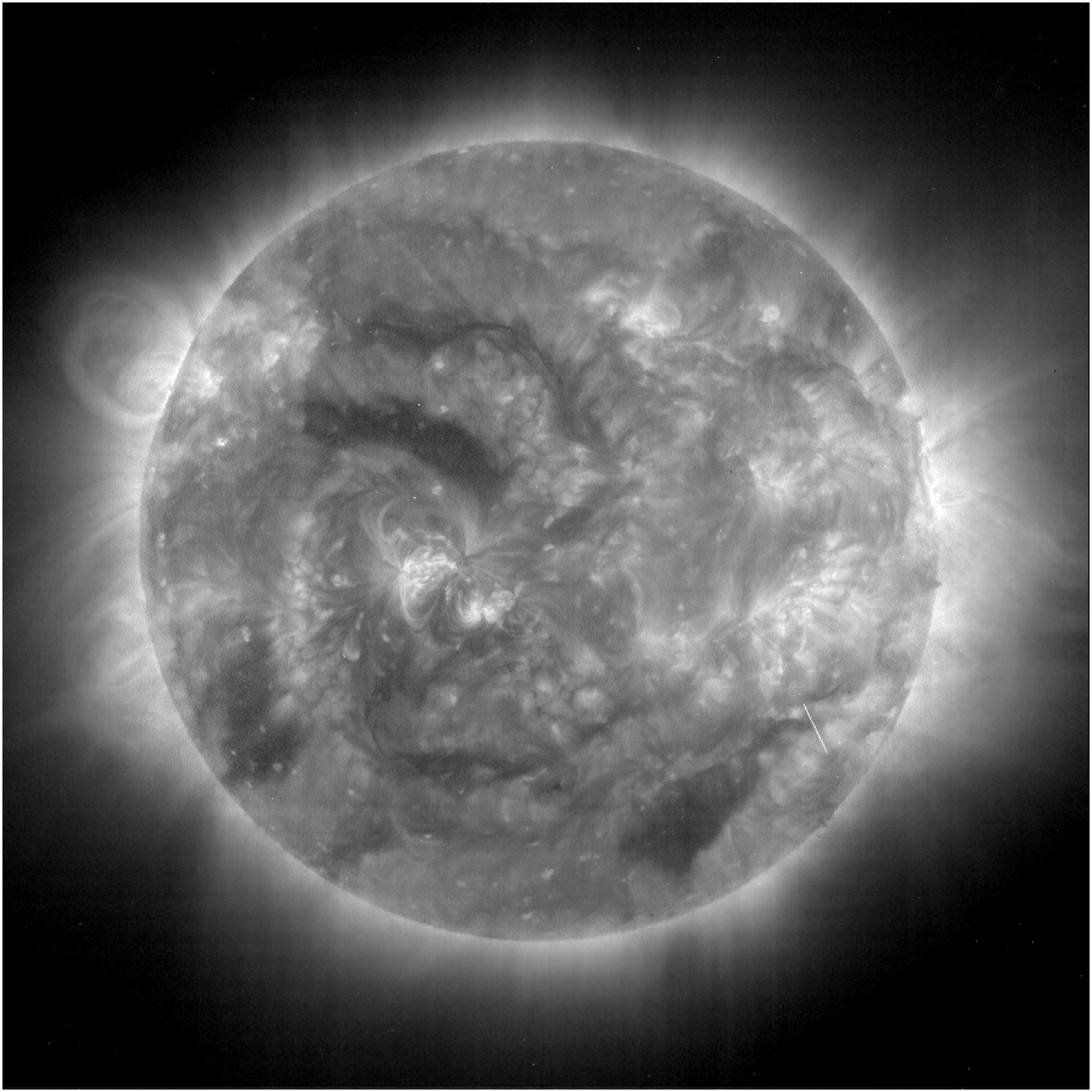}} &
      %\vector(10,10){15}
      {\includegraphics[width=18pc,height=18pc]{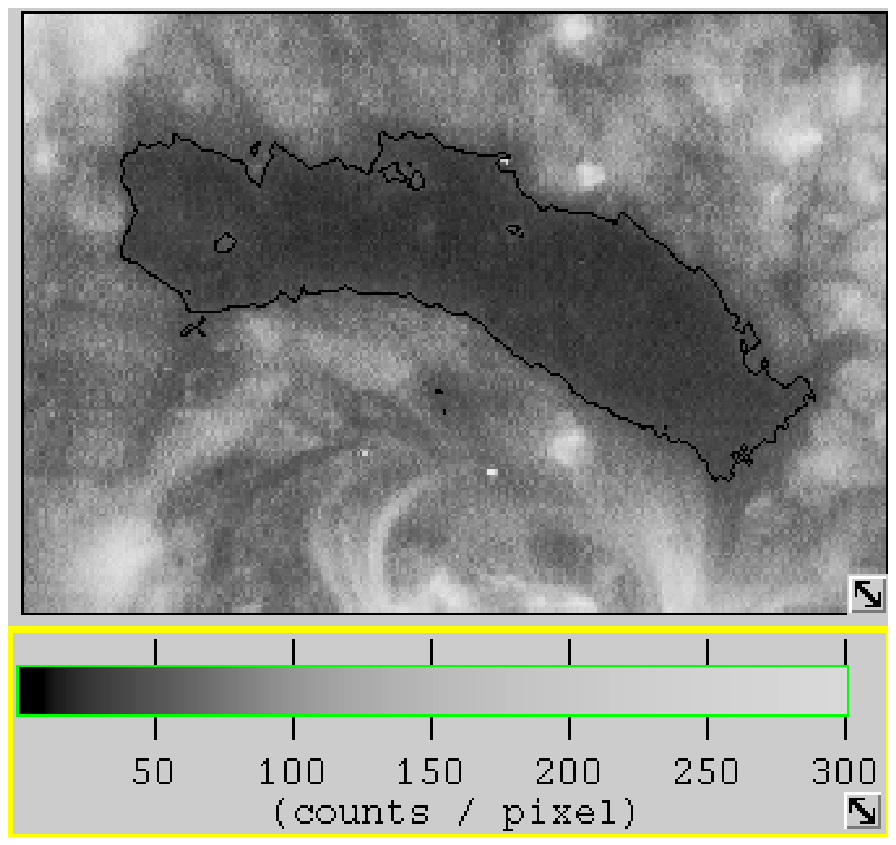}} \\
\end{tabular}
    \caption{Left side Fig 1 (a) shows full-disk SOHO/EIT 195 $\AA$ image of 01-01-2001, 00:24:11 UT with CH (in the north eastern hemisphere and close to center)
and Fig 1(b) illustrates threshold DN contour map of the same CH.}
\end{center}
\end{figure}

\begin{figure}
\begin{center}
    Fig 2(a) \hskip 40ex Fig 2(b)
    \vskip -3.5ex
    \begin{tabular}{cc}
      {\includegraphics[width=18pc,height=18pc]{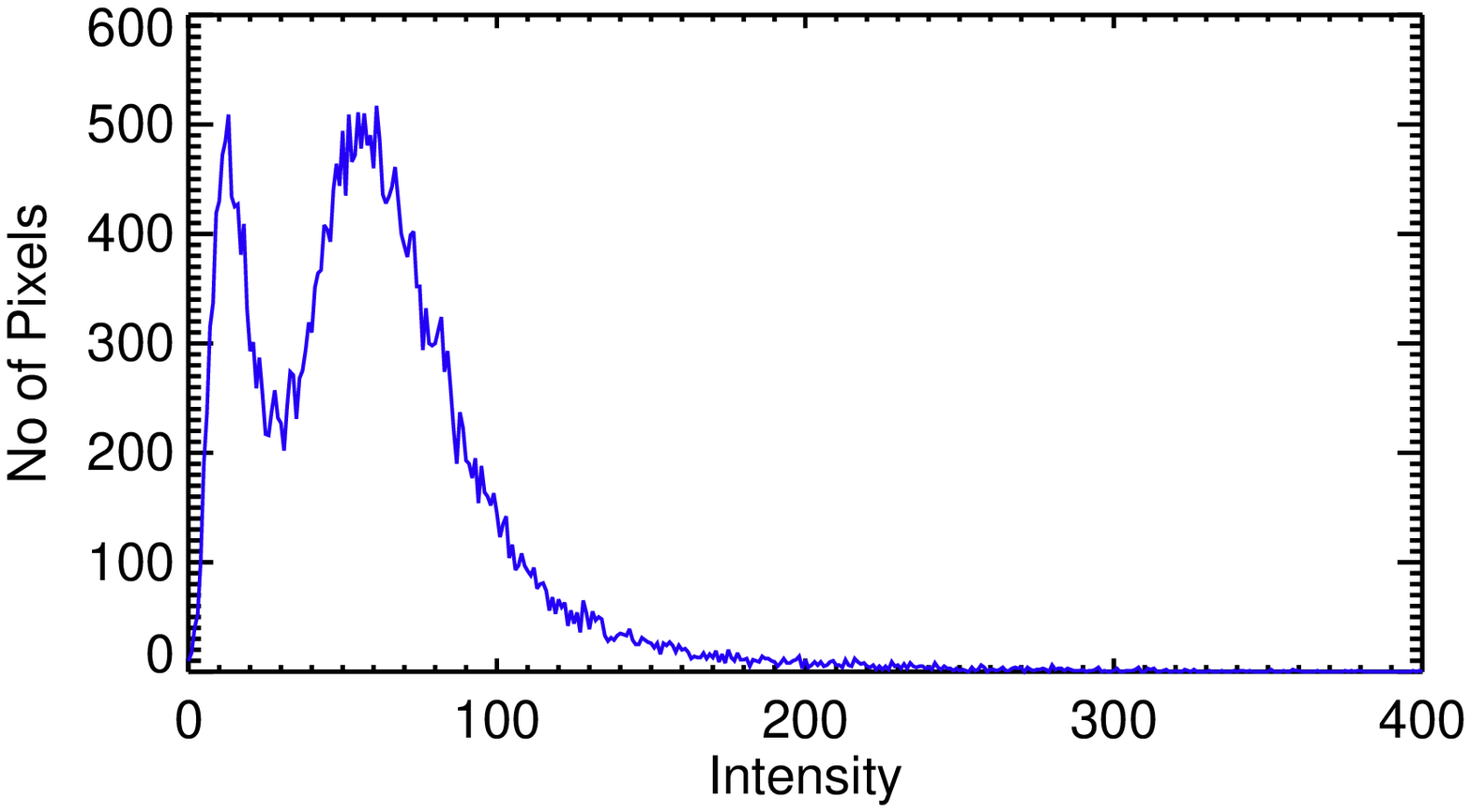}} &
      %\vector(10,10){15}
      {\includegraphics[width=18pc,height=17pc]{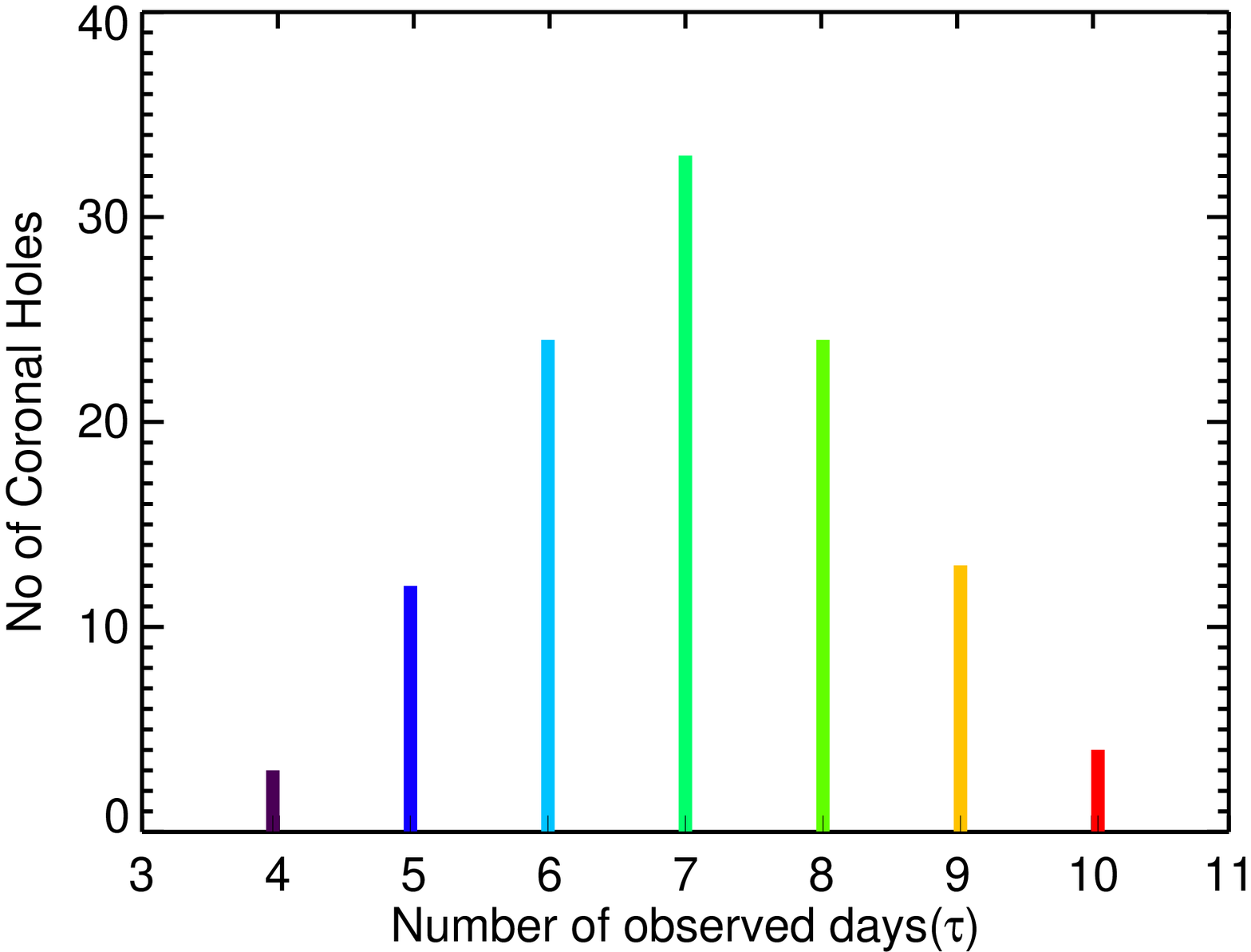}} \\
\end{tabular}

\caption{The figure on left side (Fig 2(a)) illustrates DN histogram
of a typical coronal hole. Whereas right figure (Fig 2(b)) illustrates total number of CH for 
number of days ($\tau$) observed on the solar disk.}
\end{center}
\end{figure}

Following the previous method (Hiremath 2002) of computation of
rotation rates of sunspots, daily siderial rotation rates $\Omega_{j}$ of the CH are computed as follows
\begin{equation}
\Omega_{j}=\frac{(L_{j+1} - L_{j})}{(t_{j+1} - t_{j})} + \delta \Omega \, ,
\end{equation}
where L$_{j}$, L$_{j+1}$ are average longitudes of the CH for the two consecutive days
 t$_{j}$ and t$_{j+1}$ respectively, $j=1,2, ..n-1$, $n$ is number
of days of appearance of CH on the visible solar disk and,
 $\delta \Omega$ is a correction factor for the
orbital motion of the Earth around the Sun.
Strictly speaking, this correction factor is due to orbital motion of the SOHO spacecraft
around the sun. Compared to the distance between the sun and earth, the
distance between the SOHO satellite and the earth is very small and hence
orbital distances of earth and the satellite are almost same and hence
the correction factor $\delta \Omega$ is $\sim$ 1 deg/day. For the present work, this approximation
is sufficient. However, if one wants to find the long term ($\sim$ 11 yrs)
variation of rotation rates, correction factor $\delta \Omega$
should be computed accurately (Ro\v{s}a {\em et.al.} 1995; Wittmann 1996; Braj\v{s}a {\em et al.} 2002).
From the first and second
day appearances of CH, one can compute the rotation rate $\Omega_{1}$
that we call as {\em first rotation rate}. Similarly for other
successive days, rotation rates $\Omega_{2}$, $\Omega_{3}$, {\em etc.,}
are computed.
For each computed rotation
rate of CH, the respective latitude is
assigned as the average of two latitudes corresponding to the two longitudes.
We also compute standard deviation and error bars of the average heliographic
coordinates and rotation rates. Here onwards computation of rotation rates of CH from 
equation (2) is called as {\em First Method}.  

\section{RESULTS}

\begin{figure}
\begin{center}
    Fig 3(a) \hskip 40ex Fig 3(b)
   \vskip -3.5ex
    \begin{tabular}{cc}
      {\includegraphics[width=18pc,height=18pc]{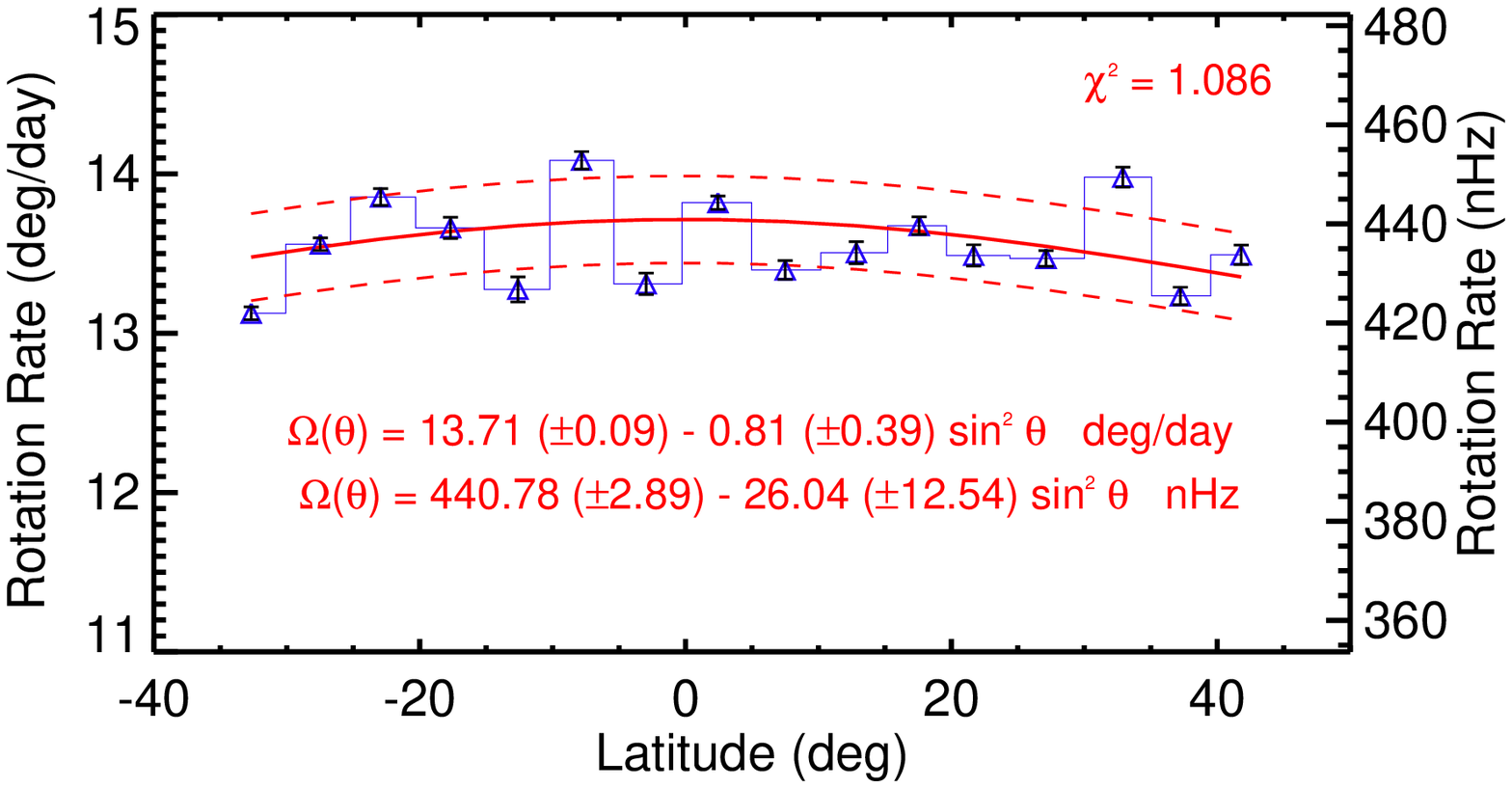}} &
      %\vector(10,10){15}
      {\includegraphics[width=18pc,height=18pc]{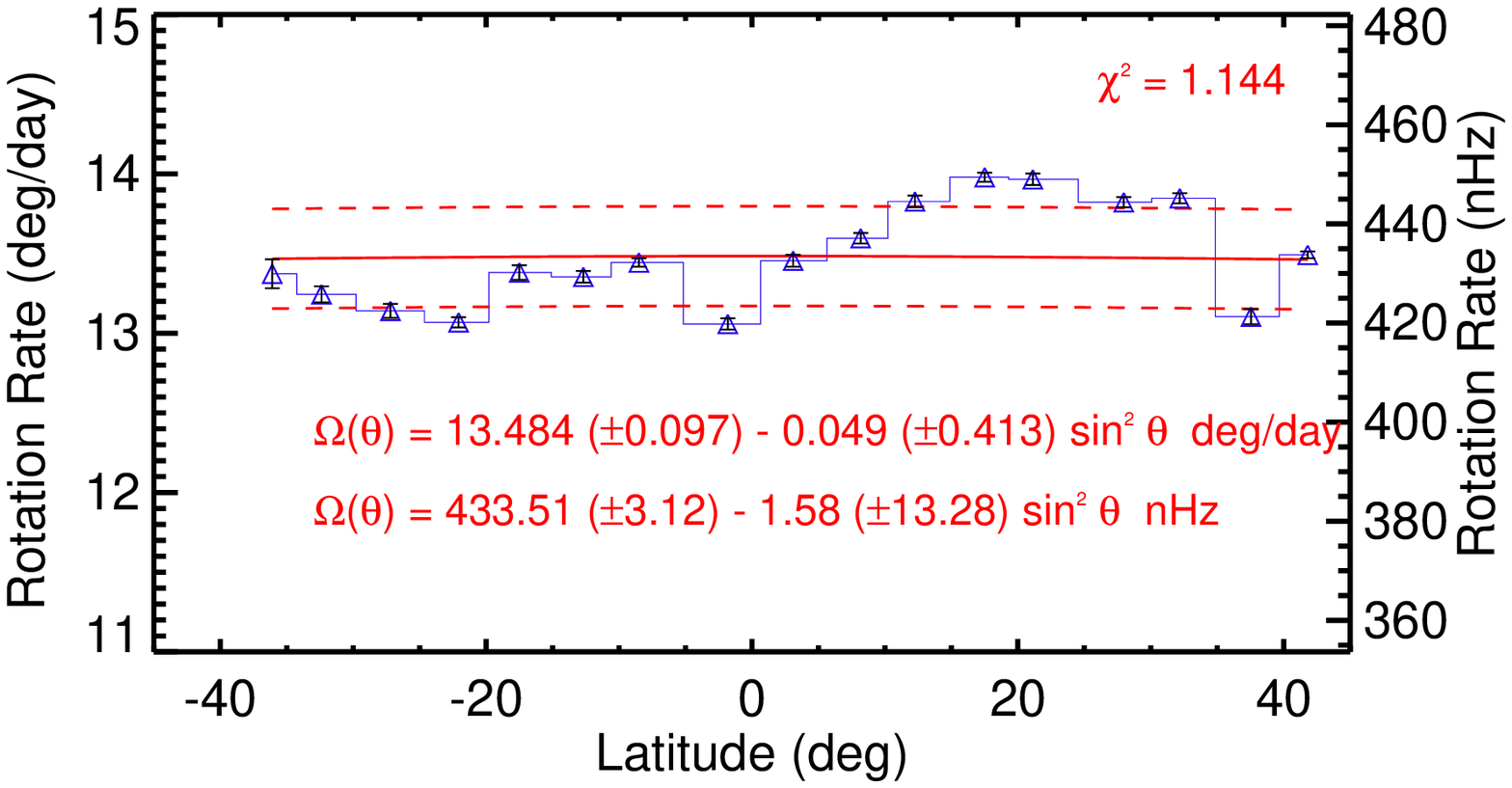}} \\
\end{tabular} 
\caption{For different latitudes, rotation rates of coronal holes
computed from the first method (see section 2).
Fig 3(a) for the rotation rates of coronal holes that occur between 
65 degrees east and west of the central meridian distance and Fig 3(b)
illustrates the rotation rates of CH that occur between 45 degrees
east and west of the central meridian distance. In both the figures blue bar plot 
represents the observed rotation rates; red dashed lines represent
the one standard deviation (that is computed from all the data points)
error bands and, the red continuous line represents a least-square fit
 of the form $\Omega (\theta)=\Omega_{0}+\Omega_{d} sin^{2} \theta$ to the observed values.
 $\Omega (\theta)$ is the observed CH rotation rate,
$\theta$ is the latitude, and $\Omega_{0}$ and $\Omega_{d}$ are the constant coefficients
determined from the least square fit. $\chi^{2}$ is a measure of goodness of fit. }
\end{center}
\end{figure}

\begin{figure}
\begin{center}
    Fig 4(a) \hskip 40ex Fig 4(b)
   \vskip -3.5ex
    \begin{tabular}{cc}
      {\includegraphics[width=18pc,height=18pc]{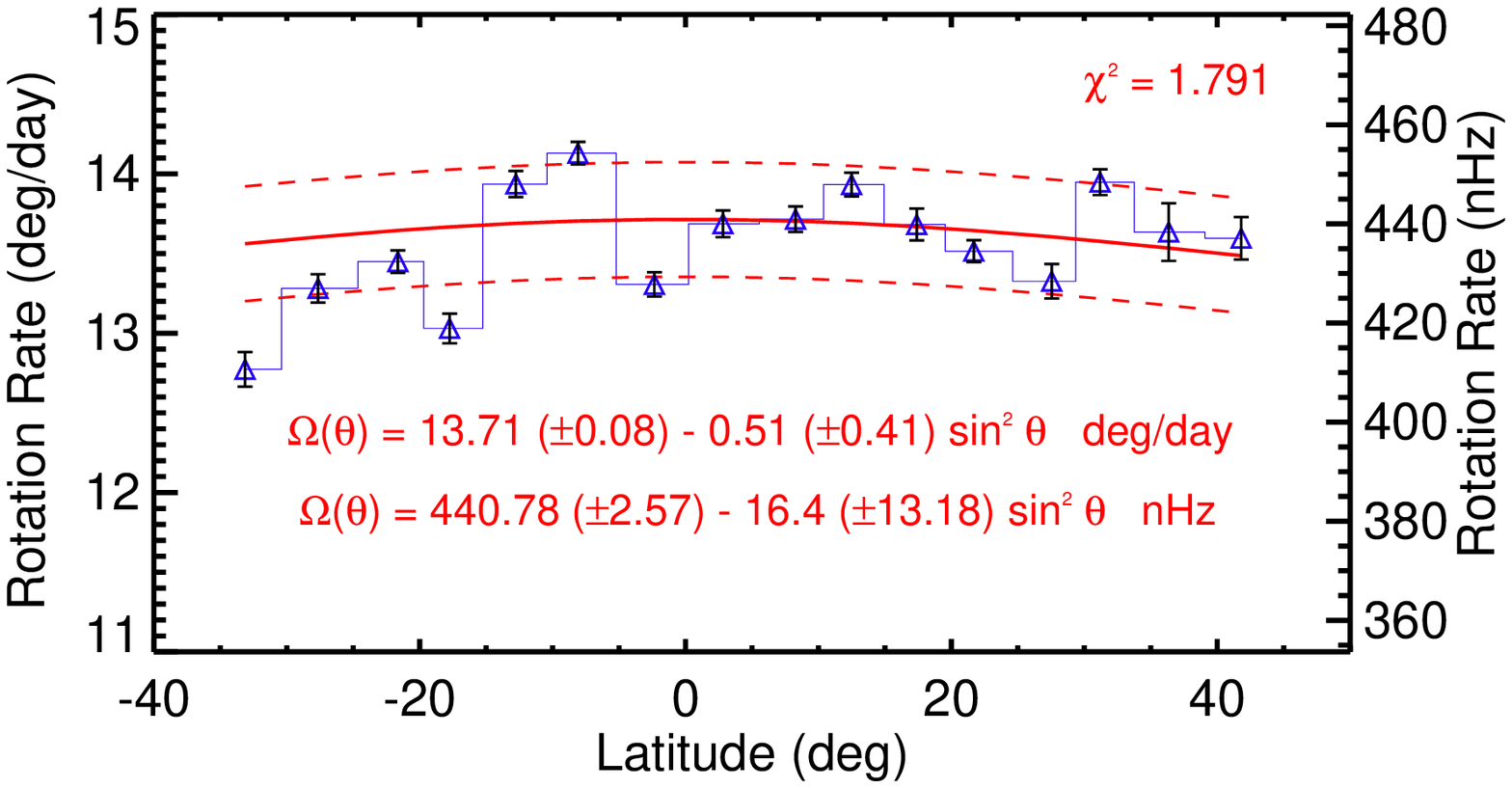}} &
      %\vector(10,10){15}
      {\includegraphics[width=18pc,height=18pc]{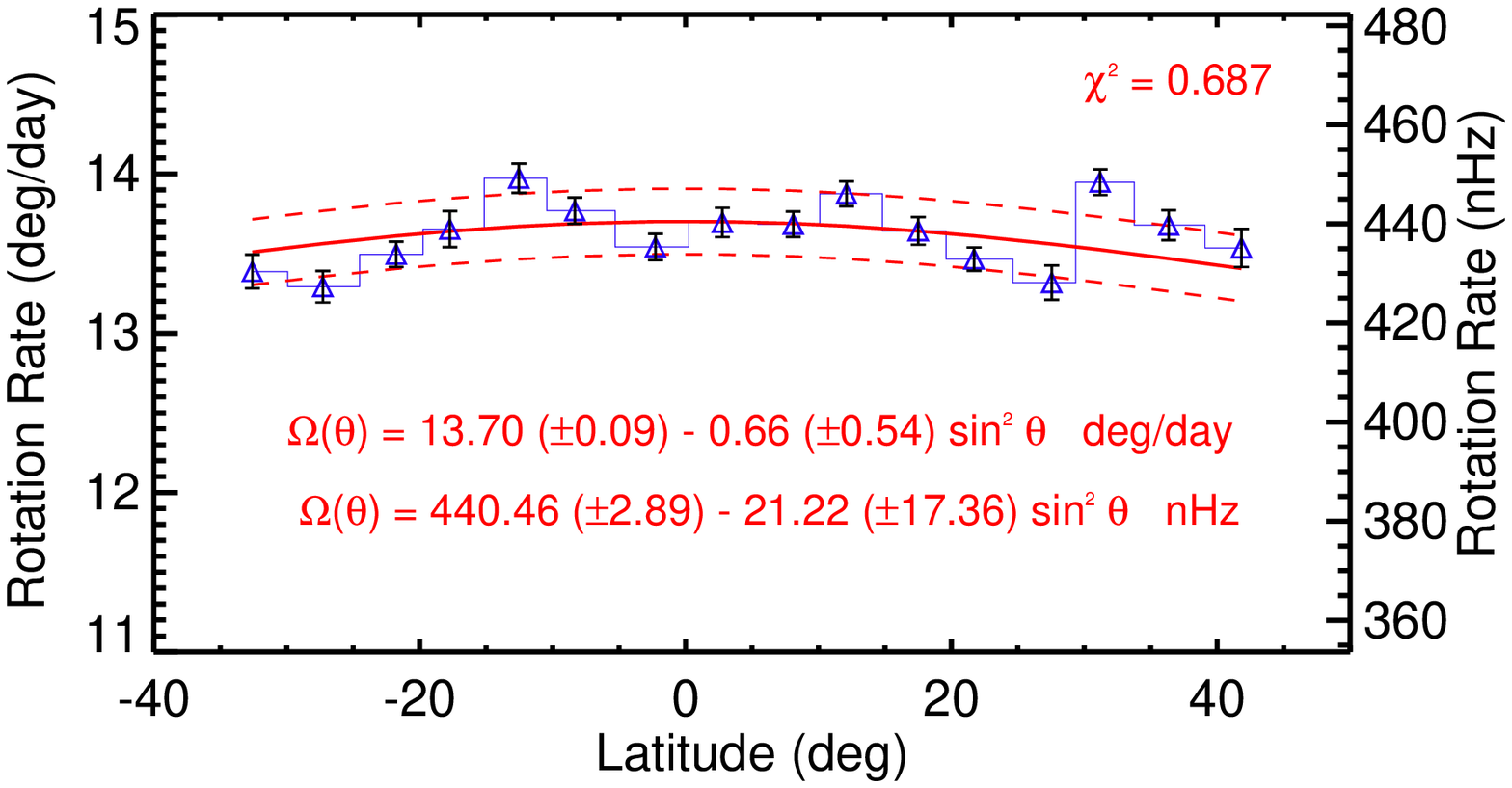}} \\
\end{tabular}
\caption{For different latitudes, rotation rates of coronal holes
computed from the second method (see section 3.1).
Fig 4(a) illustrates the rotation rates of coronal holes that occur between
65 degrees east and west of the central meridian distance and, Fig 4(b)
illustrates the rotation rates of CH that occur between 45 degrees
east and west of the central meridian distance. In both the figures blue bar plot
represents the observed rotation rates; red dashed lines represent
one standard deviation (that is computed from all the data points)
error bands and, the red continuous line represents a least-square fit
 of the form $\Omega (\theta)=\Omega_{0}+\Omega_{d} sin^{2} \theta$ to the observed values.
 $\Omega (\theta)$ is the observed CH rotation rate,
$\theta$ is the latitude, and $\Omega_{0}$ and $\Omega_{d}$ are the constant coefficients
determined from the least square fit. $\chi^{2}$ is a measure of goodness of fit.}
\end{center}
\end{figure}

We followed the following criteria in selecting CH data:
(i) In order to avoid projection effects (especially
coronal holes near both the eastern and the western limbs), we considered only
the coronal holes that emerge within 65$^\circ$  central meridian distance,
(ii) the coronal hole must be compact, independent, not elongated in latitude,
and, (iii) during its passage across the solar disk it should not merge with other coronal holes.
For the period of observations from 2001 to 2008, a total of 113 CH satisfy these criteria.
 We define the term $\tau$ of a CH as total number of
days observed on same part of the solar disk satisfying the afore mentioned criteria.
%although {\em actual life span} of CH must be higher.
%It is to be noted that the definition, {\em apparent life span}, of CH is a relative value
%that is computed from the visible solar disk and absolute value must be different. 
Suppose we assume that CH decay due to magnetic diffusion only,
as the dimension $L$ of CH is very large (from the following section 3.1, one
can note that area $A$ is $\sim 10^{20} cm^{2}$), magnetic diffusion time scale
$\tau$ (${L^{2} \over \eta}$ $\sim$ ${A \over \pi \eta}$, where $\eta$
is magnetic diffusivity and area $A$ of CH is assumed
to be a circle; magnetic diffusivity in the corona is
considered to be $\sim$ $10^{13} cm^{2}sec^{-1}$ (Krista 2011; Krista {\em et.al} 2011)) is estimated to
be $\sim$ 2 months. Hence, there is a possibility that CH might
have reappeared again on the visible disk and
might have diffused in the solar atmosphere. Hence, {\em actual life span} of CH must be 
of longer duration.
In Fig 2(b), for different $\tau$, we present occurrence number of CH considered for this study.  

During their evolutionary passage over the solar disk, we compute rotation rates 
and assign respective latitudes. If the CH exists for $n$ days, then its
$\tau$ is $n$ days and, total number of rotation rates is $(n-1)$.
Rotation rates of non-recurrent CH that appear and disappear on the 
visible disk are computed. According to above definition, and in the present data 
set (see Fig 2(b)), we find 4 CH that appear for 10 days, 13 CH for 9 days and so on. 
Integrated over all latitudes and in both the hemispheres, we determined a total of 683 rotation rates.
\subsection{Average Rotation Rates : Variations With Respect to Latitude and Area}
During their passage over the solar visible disk,
daily rotation rates of CH are computed.
In both the hemispheres, for each latitude bin of 5$^\circ$, we collect rotation rates and
compute average rotation rates with their respective standard deviations $\sigma$
and the errors (${\sigma \over \sqrt{N}}$, where N is number of rotation rates). 
%We call these average rotation rates as {\em overall rotation rates} to
%distinguish from the {\em initial rotation rates} as defined in the
%previous section. 
We present the results in Fig 3 (a) that illustrate the variation of average rotation rates of
the coronal holes for different latitudes. To be on the safer side
from the projectional effects, we also compute {\em average rotation rates} of coronal holes 
that emerge within 45$^\circ$ central meridian distances and are illustrated in Fig 3(b).
For the sake of comparison with helioseismic inferred rotation rates, in both the plots,
we include a frequency scale on the right hand side
of the vertical axis.
For different latitude bins, observed rotation rates are subjected to a least square fit of the form
$\Omega (\theta)=\Omega_{0}+\Omega_{d} sin^{2} \theta$ 
(where $\theta$ is latitude, $\Omega_{0}$ \& $\Omega_{d}$ are constant coefficients to be determined). 

There is every possibility that as the errors in determination of centers
of CH propagate to the rotation rates and hence rotation rates determined 
from the first method effectively enhance the error in the
second coefficient ($\Omega_{d}$) yielding rigid body rotation
rates of CH. Moreover, drawback of the first method is
also reflected in Fig 3(b) where unlikely asymmetrical rotation profile
in both the hemisphere is obtained. In order to minimize such
propagating errors in the rotation rates of CH determined
by the first method, we compute rotation
rates of CH in the following way and define as a {\em Second Method}.
In this method, as suggested by the referee, we fit all
the daily centroid positions of the individual coronal holes
by a first degree polynomial and, computed second coefficient (slope)
represents the rotation rate.
For each computed rotation rate of CH, the respective
latitude is assigned by averaging all the latitudes
of CH during its passage. As described in the previous paragraph we binned
the rotation rates, computed the average rotation
rates, standard deviations and error bars respectively. For
different latitude bins, average rotation rates are subjected
to a linear least square fit and the results are presented in Fig 4.  
From both the rotation laws, compared to the first coefficient, magnitude of small second
coefficient ($\Omega_{d}=-0.81(\pm 1.58)$ in Fig 3(a)
or $\Omega_{d}=-0.51(\pm 1.64)$ in Fig 4(a)) suggests 
that {\em CH rotate rigidly}.

\begin{figure}
\begin{center}
    Fig 5(a) \hskip 40ex Fig 5(b)
   \vskip -3.5ex
    \begin{tabular}{cc}
      {\includegraphics[width=18pc,height=18pc]{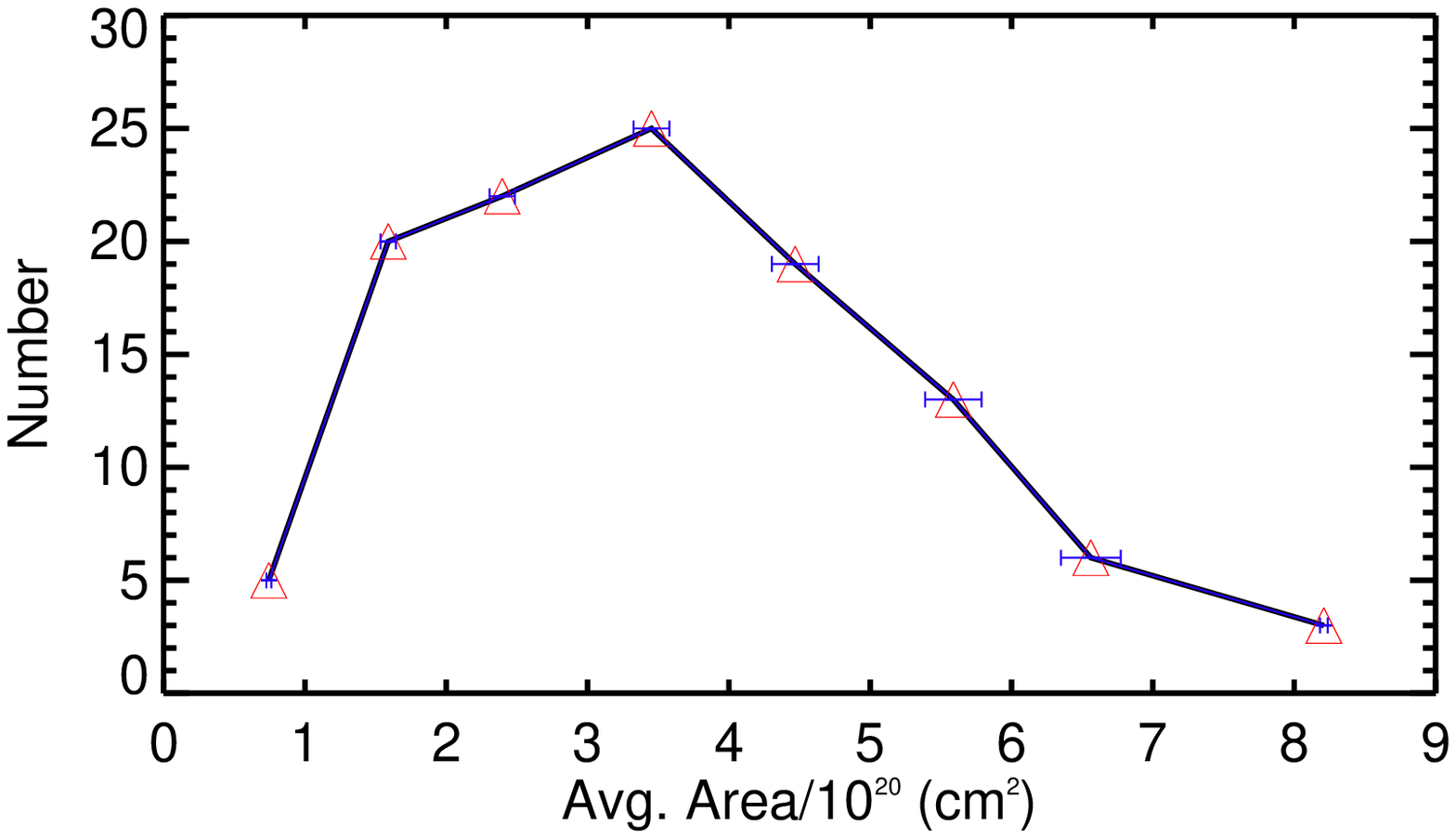}} &
      %\vector(10,10){15}
      {\includegraphics[width=18pc,height=18pc]{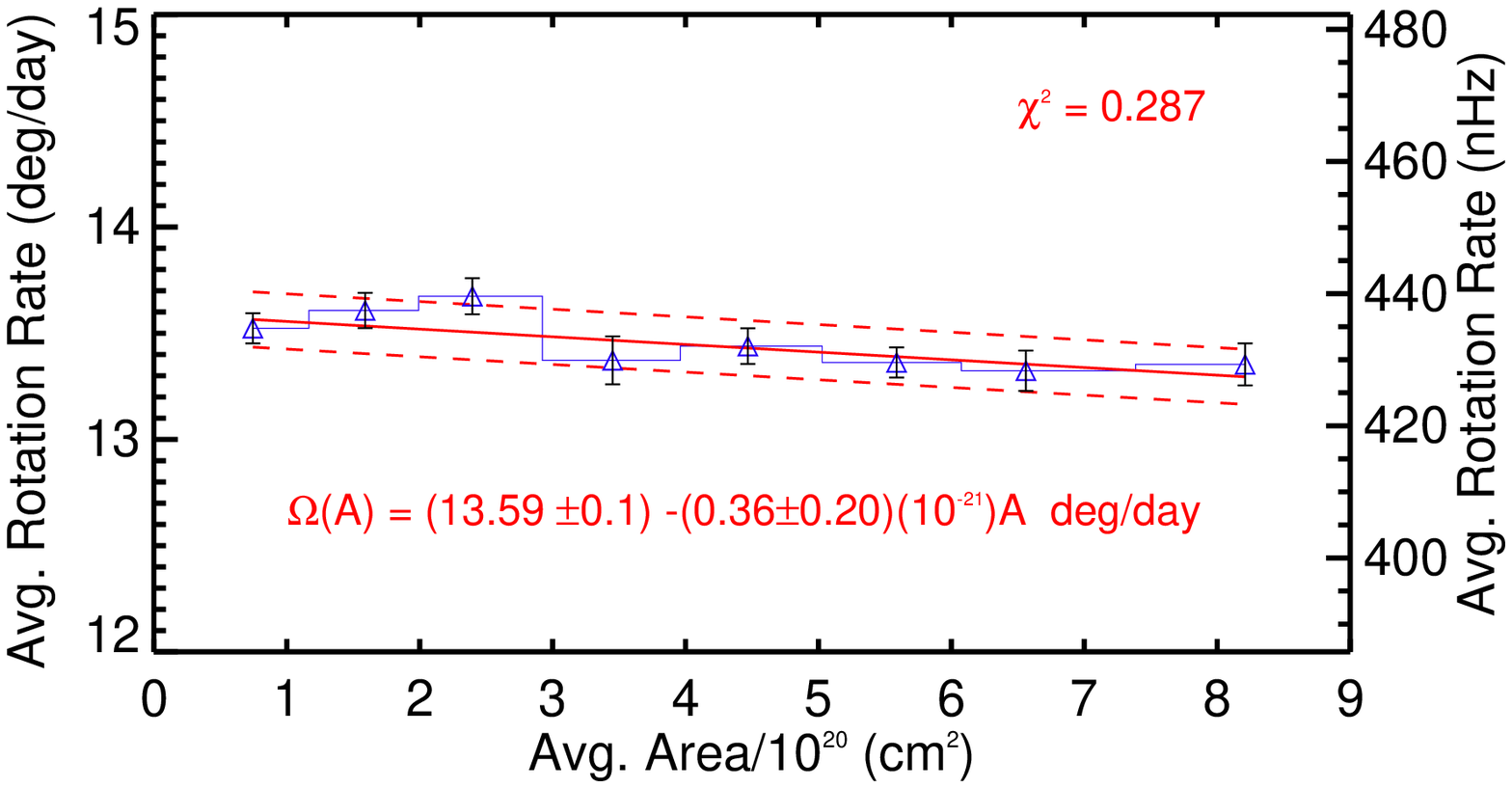}} \\
\end{tabular}
\caption{Irrespective of their latitude and number of days ($\tau$) 
observed on the disk, for different area bins, left
figure (5(a)) illustrates occurrence number of CH considered for the analysis
and right figure (5(b)) illustrates  variation of rotation rates 
for different average areas. In Fig 5(b), the blue bar plot  
represents computed rotation rates and
the red continuous line represents a least-square fit
 $Y=a+bX$ to the observed values. $Y$ is observed rotation rate of CH,
$X$ is the average area, and $a$ and $b$ are the constant coefficients
determined from the least square fit. Red dashed lines in Fig 5(b) represent
one standard deviation (that is computed from all the data points)
error bands. $\chi^{2}$ is a measure of goodness of fit.
}
\end{center}
\end{figure}

\begin{figure}
\begin{center}
    Fig 6(a) \hskip 40ex Fig 6(b)
   \vskip -3.5ex
    \begin{tabular}{cc}
      {\includegraphics[width=18pc,height=18pc]{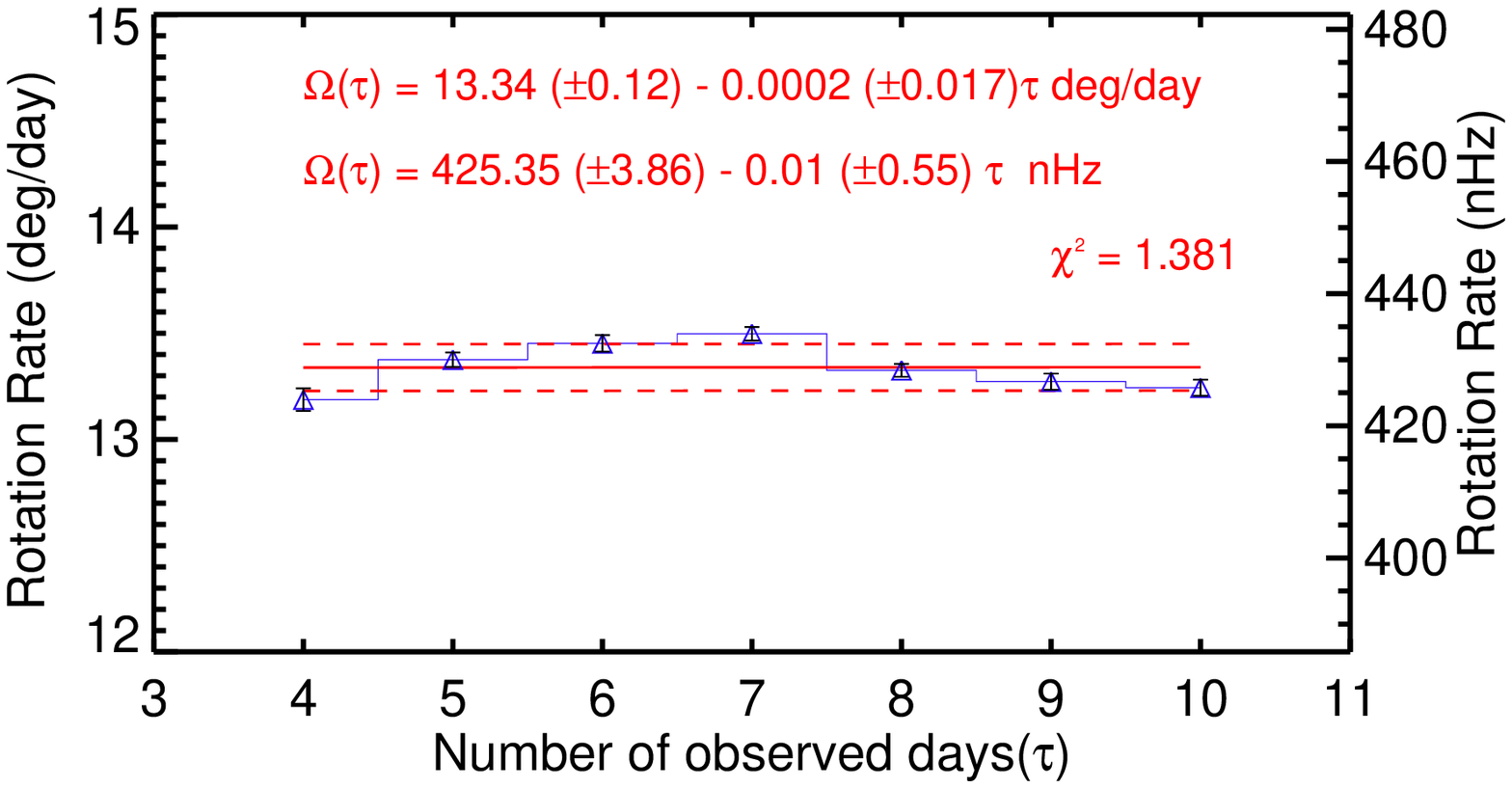}} &
      %\vector(10,10){15}
      {\includegraphics[width=18pc,height=18pc]{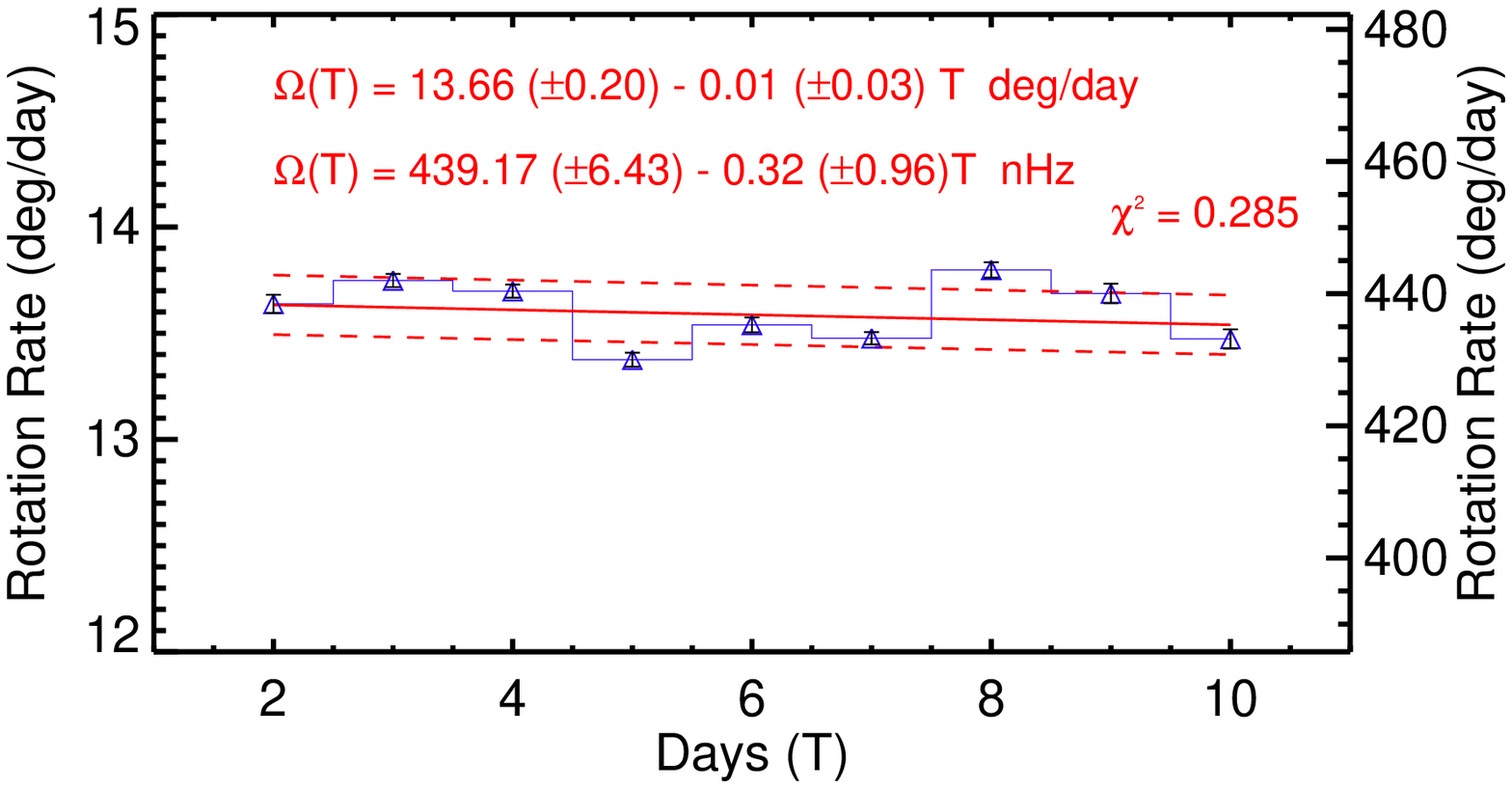}} \\
\end{tabular}
\caption{Irrespective of their area and latitude, 
left (6(a)) and right (6(b)) figures illustrate the variation of rotation rates of CH
for different $\tau$ and for different days during
their evolution respectively. In both the figures, blue bar 
plot  represents the computed rotation rates and
 red continuous line represents a least-square fit
 $Y=a+bX$ to the observed values. $Y$ is the observed rotation rate of CH,
$X$ is either $\tau$ or different days represented by $T$, and $a$ and $b$ are the constant coefficients
determined from the least square fit. Red dashed lines represent
one standard deviation (that is computed from all the data points)
error bands. $\chi^{2}$ is a measure of goodness of fit.}
\end{center}
\end{figure}

\begin{figure}
\begin{center}
 Fig 7(a) \hskip 40ex Fig 7(b)
   \vskip -2.5ex
\begin{tabular}{cc}
      {\includegraphics[width=18pc,height=18pc]{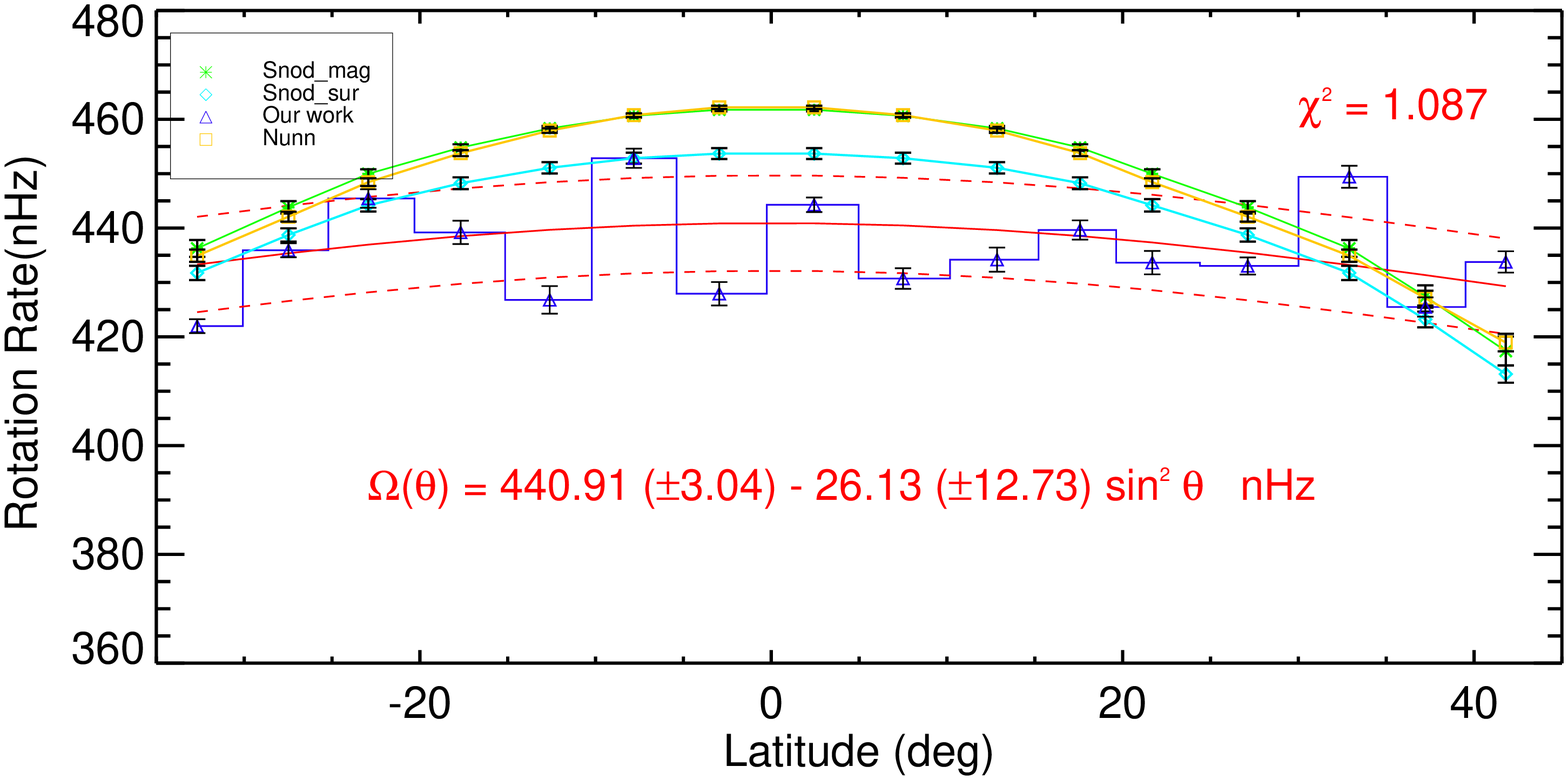}} &
      {\includegraphics[width=18pc,height=18pc]{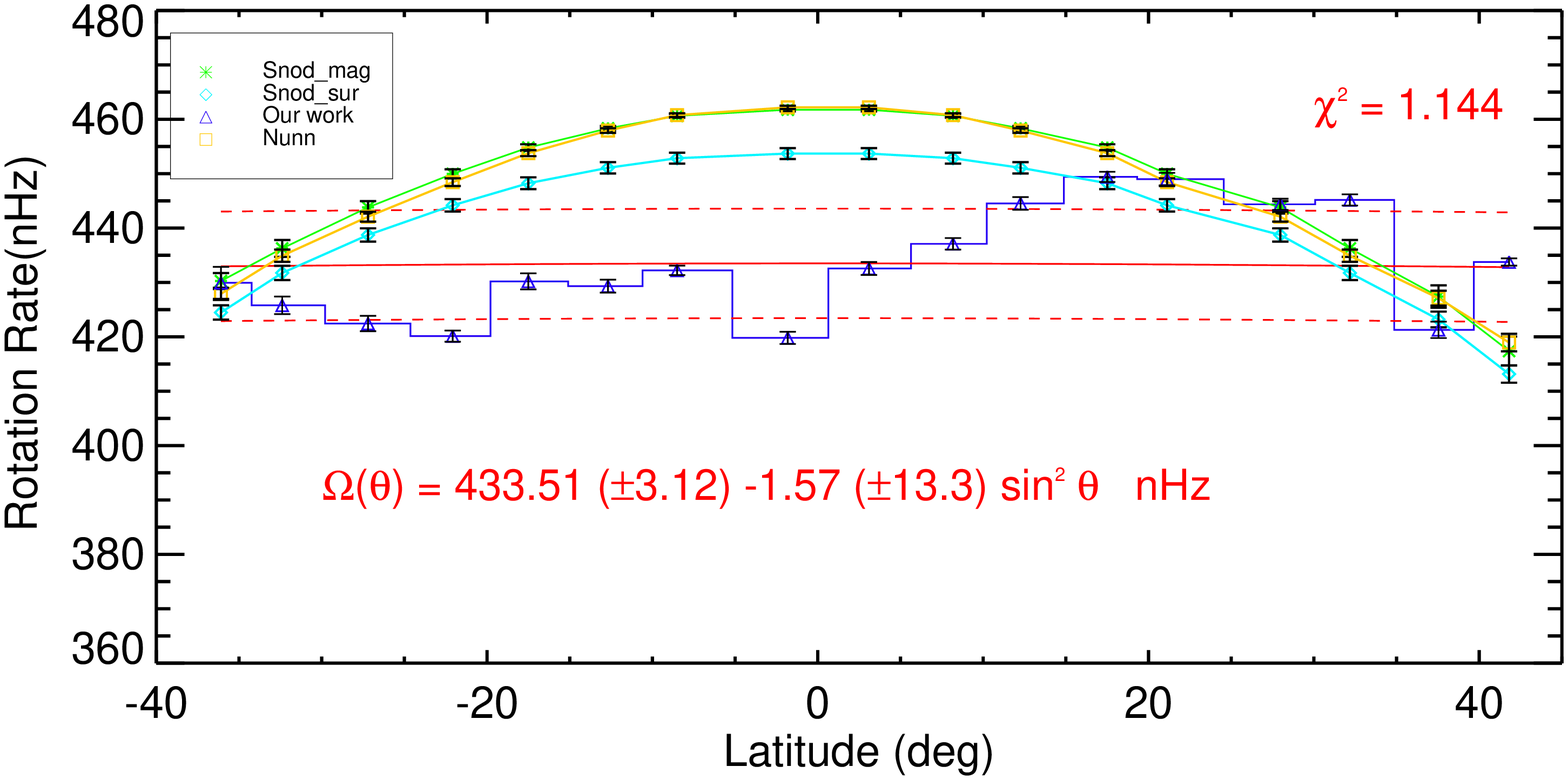}} \\
\end{tabular}
\end{center}
\caption{Irrespective of their areas, for different latitudes, 
blue bar plots that are connected by triangles in both the plots
represent the rotation rates of coronal holes
as determined from the first method (see section 2). 
Blue bar plots in both the illustrations represent rotation rates of CH that occur between
East and West of 65 (Fig 7(a)) and 45 (Fig 7(b)) degrees central meridian distances respectively.
Rotation rates of sunspots (yellow curve; Newton and Nunn 1951), magnetic activity (green
curve; Snodgrass 1983) and surface rotation (cyan curve; Snodgrass 1992)
are also over plotted. Red dashed lines in both the figures represent
one standard deviation (that is computed from all the data points)
error bands. $\chi^{2}$ is a measure of goodness of fit.}
\end{figure}

\begin{figure}
\begin{center}
 Fig 8(a) \hskip 40ex Fig 8(b)
   \vskip -2.5ex
\begin{tabular}{cc}
      {\includegraphics[width=18pc,height=18pc]{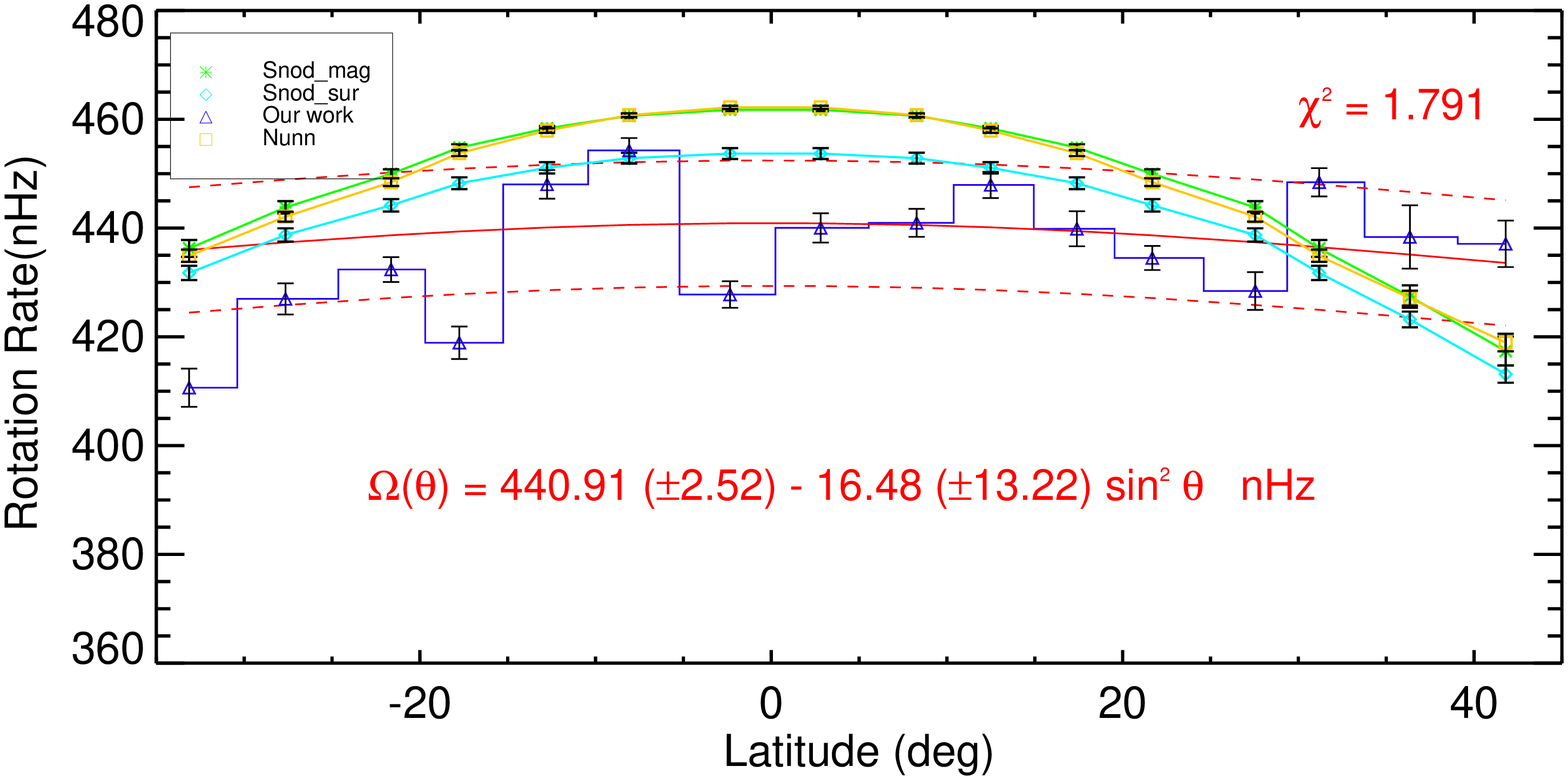}} &
      {\includegraphics[width=18pc,height=18pc]{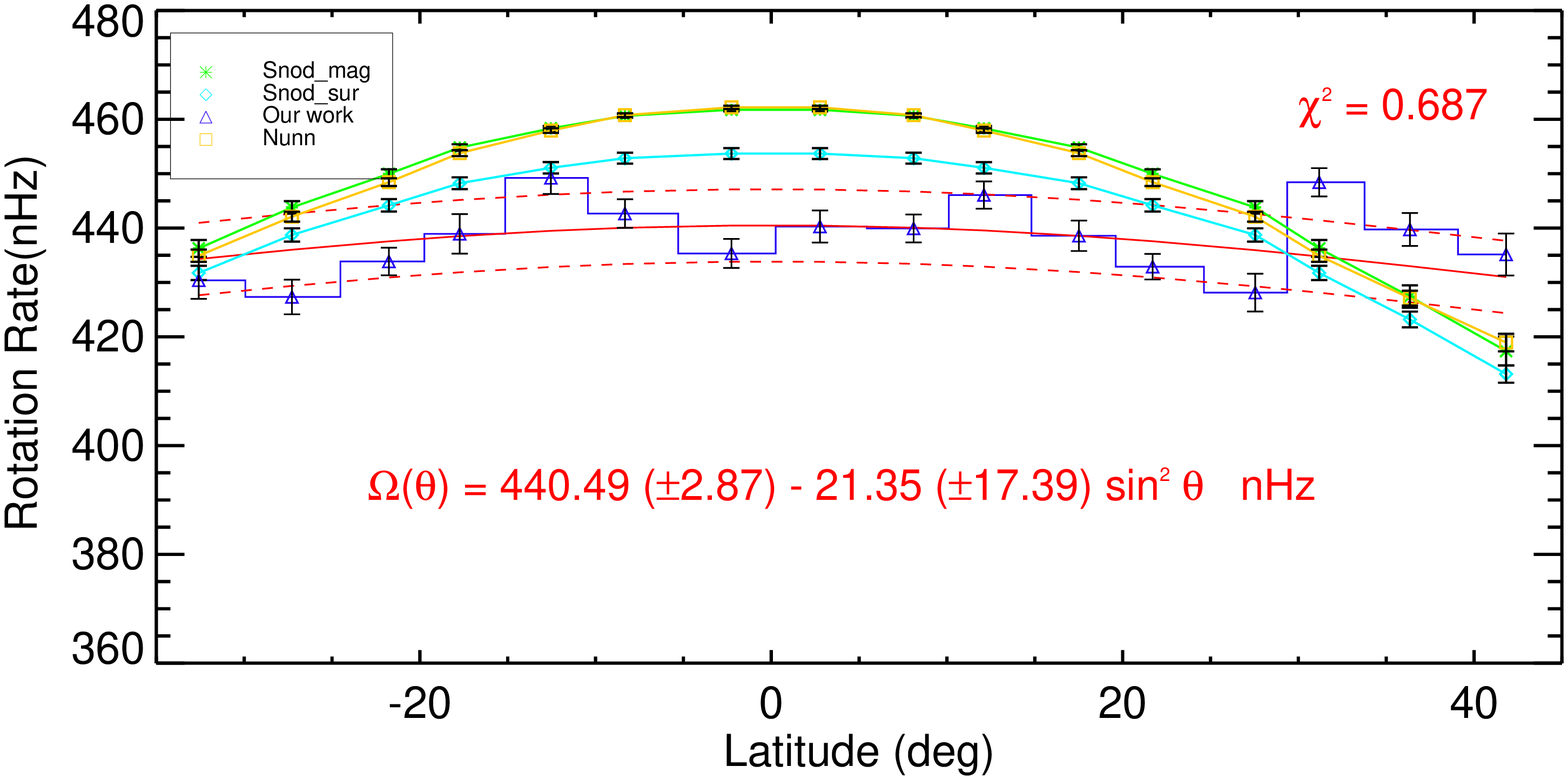}} \\
\end{tabular}
\end{center}
\caption{
Irrespective of their areas, for different latitudes,
blue bar plots that are connected by triangles in both the plots
represent the rotation rates of coronal holes
as determined from the second method (see section 3.1).
Blue bar plots in both the illustrations represent rotation rates of CH that occur between
East and West of 65 (Fig 8(a)) and 45 (Fig 8(b)) degrees central meridian distances respectively.
Rotation rates of sunspots (yellow curve; Newton and Nunn 1951), magnetic activity (green
curve; Snodgrass 1983) and surface rotation (cyan curve; Snodgrass 1992)
are also over plotted. Red dashed lines in both the figures represent
one standard deviation (that is computed from all the data points)
error bands. $\chi^{2}$ is a measure of goodness of fit.
}
\end{figure}

As sunspots show different rotation rates
for the small and big areas (Hiremath 2002),
it is interesting to know whether similar
variations in rotation rates exist in case of coronal holes.
As CH evolve, their area also changes and question
arises: for which area during the evolutionary passage,
rotation rate has to be considered. 
For this purpose, we adopt the following method. Daily areas
and rotation rates of CH are computed. For all the days
of CH$^{'}$ existence, average area and rotation
rates are computed. Further, irrespective of their latitude and $\tau$, rotation rates are collected 
for the area bins (0-1)$\times 10^{20} cm^{2}$, 
(1-2)$\times 10^{20} cm^{2}$, etc, and {\em mean rotation rates} are
computed. In Fig 5(a), we present occurrence number of CH for 
 different area bins. Whereas, irrespective of their latitudes and $\tau$,
for different area bins, Fig 5(b) illustrates the {\em mean rotation rates} of CH.
It is important to note from Fig 5(b) that, unlike
sunspots, for different areas, all the coronal holes (as the second coefficient
 is almost zero, i.e., $(0.36\pm 0.20)\times 10^{-21}$)
{\em rotate rigidly}. This important result
implies that all the CH must originate from same region of the solar
interior that rotate rigidly.  

\subsection{Average Rotation Rates : Variations With
Respect to $\tau$ and Daily Evolution}
In order to check dependency of rotation rates of CH 
with respect to number of observed days $\tau$, daily rotation rates are computed
during their evolution. As described in section 3, if CH has $\tau$
of $n$ days, we have $(n-1)$ rotation rates. Irrespective
of their areas and the  latitude, for each $\tau$,
rotation rates are collected and average rotation rate is
computed and the results are illustrated in Fig 6(a).
We find that, rotation rates of CH are independent of $\tau$. 

Further, irrespective of their area and $\tau$, 
we combined daily rotation rates 
for all the latitudinal bins; we present the resulting daily average rotation rates
in Fig 6(b). If coronal holes rotate rigidly and are independent of latitude, then the
integrated rotation rates for all the latitudes should remain constant. For example,
let us consider the rotation law (red continuous line) over plotted on Fig 6(b).
From this law, when one computes the difference between
rotation rates of the first day and the 10th day, the difference is found
to be $\sim$ 0.1 degree/day, approximately same magnitude as the formal uncertainty
in the value for each bin, once again strongly suggesting that, {\em for all the 
days during their evolutionary passage, coronal holes rotate
rigidly}.

\begin{table*}
 \centering
% \begin{minipage}{220mm}
 \begin{minipage}{180mm}
\caption{Sidereal rotation rates (deg/day) obtained by the present and previous studies.}
%\hskip-10ex
%\hskip-5ex
 \begin{tabular}{@{}lllllll@{}}
 \hline
Different & Observations & Wavelength &  \multicolumn{2}{c}{Coefficients}    &{$|\Omega_{d}$}/{$\Omega_{0}|$}  & $\chi^{2}$\\
regions &  &   region & $\Omega_{0}$     &   $\Omega_{d}$        &       &\\
 \hline
Corona       & Coronal holes$^{1}$ &  EUV     & 13.71     &-0.81      &0.059   &1.086\\
Corona        & Coronal holes$^{1a}$  &  EUV     & 13.48                 &-0.05      &0.004  &1.144\\
%Corona            & Coronal holes$^{1b}$            &  EUV     & 13.07                 &  0.08     &0.006    &\\
Corona       & Coronal holes$^{2}$ &  EUV     & 13.71                 &-0.51      &0.037   &1.791\\
Corona        & Coronal holes$^{2a}$            &  EUV     & 13.70                 &-0.66      &0.048   &0.686\\ 
Corona        & Coronal holes$^{3}$     &  EUV     & 13.61                 &-0.15      &0.011   &0.202\\
Photosphere       & Doppler Shift$^{4}$            & Visible  & 14.11                &-1.70      &0.121    &\\
Photosphere       & Surface magnetic$^{5}$         & Visible  & 14.37                &-2.30      &0.160    &\\
Photosphere       & Sunspots$^{6}$                 & Visible & 14.38                &-2.96     &  0.206    &\\
Photosphere       & Sunspots$^{7}$                 & Visible & 14.37                &-2.59     &  0.180    &\\
Radiative Core    & Helioseismic$^{8}$   &          & 13.63                &-0.64     &0.047    &5.401\\
\hline
\end{tabular}
\end{minipage}
$^1${Average rotation rates from the First Method and for the CMD (+65$^\circ$ to -65$^\circ$)}

$^1a${Average rotation rates from the First Method and for the CMD (+45$^\circ$ to -45$^\circ$)}
%\tablenotetext{1b}{Initial rotation rates from the First Method and for the CMD (+45$^\circ$ to -45$^\circ$)}

$^2${Average rotation rates from the Second Method and for the CMD (+65$^\circ$ to -65$^\circ$)}

$^2a${Average rotation rates from the Second Method and for the CMD (+45$^\circ$ to -45$^\circ$)}

$^3${First rotation rates for the CMD (+45$^\circ$ to -45$^\circ$)}

$^4${Snodgrass (1992); $^5$Snodgrass (1983); $^6$Newton \&  Nunn (1951); 

$^7$Braj\v{s}a {\em et.al.} 2002; $^8$Antia and Basu (2010)}

{*}{CMD-Central Meridian Distance}
\end{table*}

\subsection{Comparison of Rotation Rates of CH With
Other Activity Indices}
Compared to rotation rates obtained by other surface activity
indices (Figures 7 and 8), (i) coronal holes rotate almost like a rigid body
 and, (ii) on average, coronal holes rotate
slower ($\sim$ 440 nHz) than the rotation rates of
other activity indices over the latitude range $-40$ to $+40$.
The ratio $R=|{\Omega_{d}\over \Omega_{0}}|$
of the two coefficients of each rotational law
gives a sense of whether the rotation is
rigid or differential. For example, if one
computes this ratio for sunspots ($R_{sunspot}$)
and for coronal holes ($R_{coronal\_hole}$), it is clear that
$R_{sunspot}$ $\gg$ $R_{coronal\_hole}$,
as can also be seen from the fifth column of
Table 2. In this table, goodness of fit $\chi^{2}$ is also
given in the last column. Small value of $\chi^{2}$ (typically
$\chi^{2}$ should be $\le$ (N-n), where $N$ is total number of data points
and $n$ is degrees of freedom, in this case $n=2$) implies fit is very good.
 At least compared with any features lower
in the solar atmosphere, it is clear that {\em CH rotate rigidly}.

\begin{figure}
\begin{center}
Fig 9(a) \hskip 40ex Fig 9(b)
   \vskip -2.5ex
\begin{tabular}{cc}
      {\includegraphics[width=18pc,height=18pc]{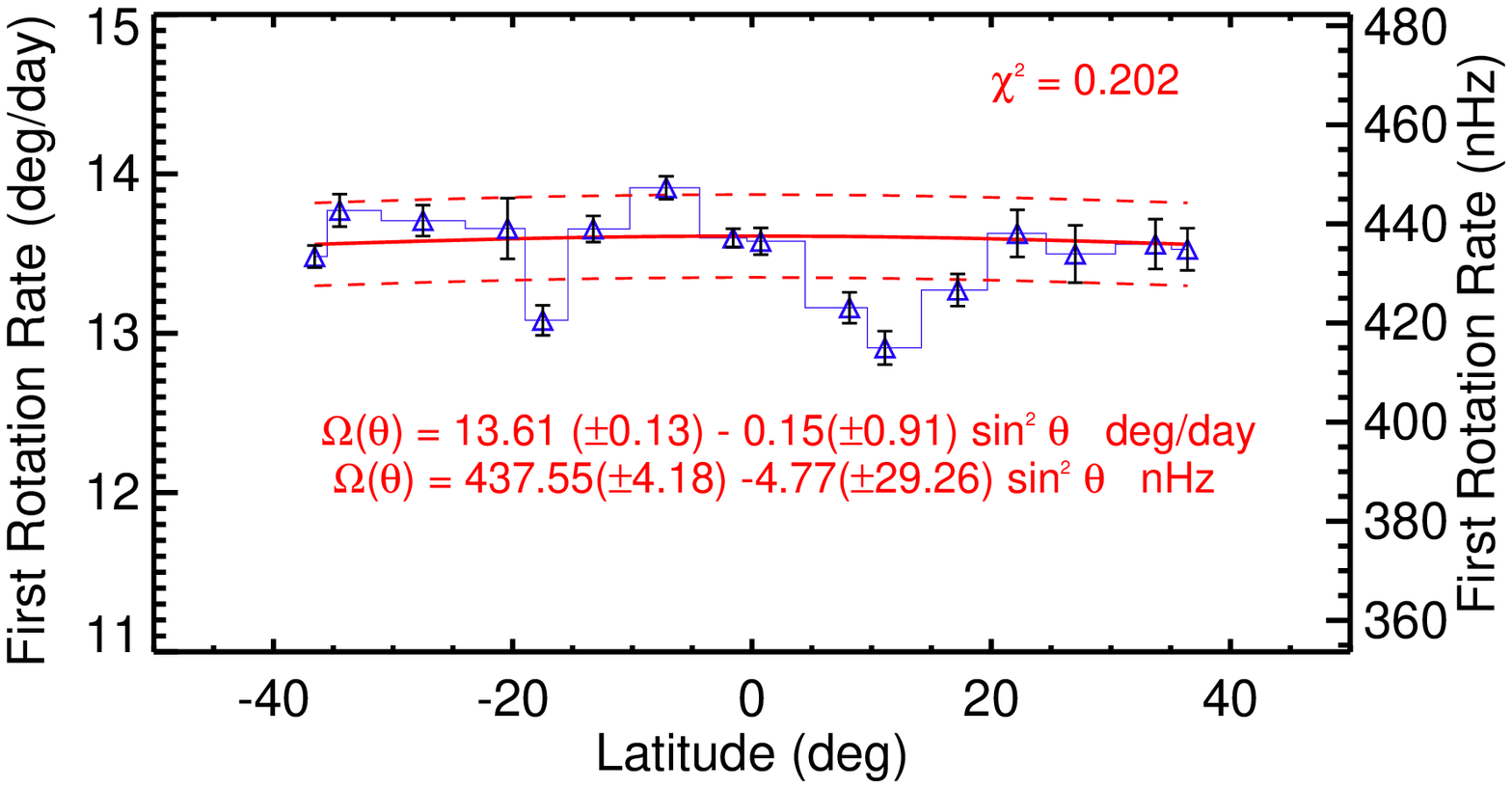}} &
      {\includegraphics[width=18pc,height=18pc]{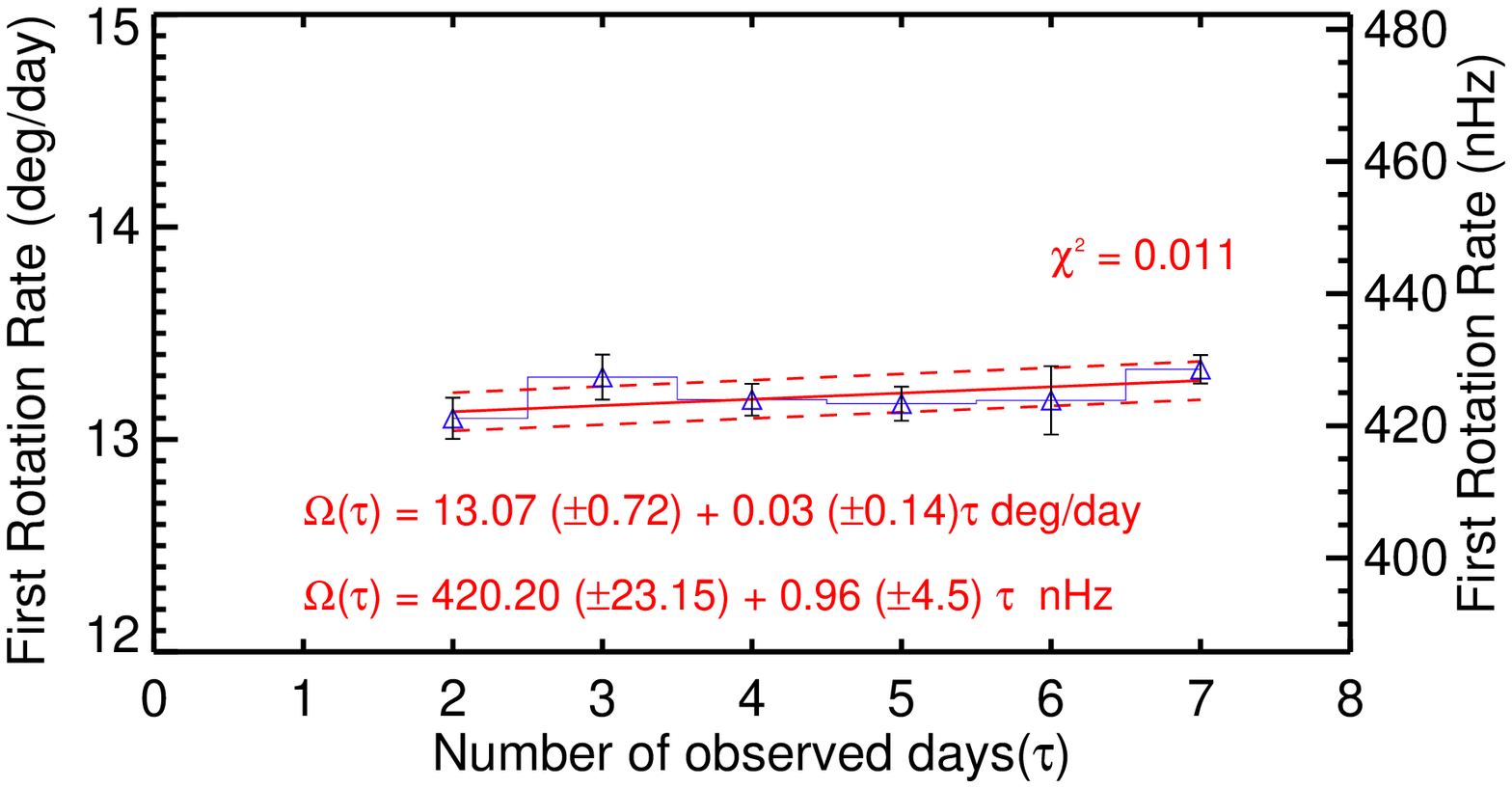}} \\
\end{tabular}
\end{center}
\caption{ Irrespective of their area and number of days ($\tau$) observed
on the disk, left figure (9a)
illustrates the variation of first rotation rates of
CH with respect to latitude. Irrespective of their
area and latitude, for different $\tau$, right figure (9b) illustrates the variation
of first rotation rates of CH. Red dashed lines in both the figures represent
one standard deviation (that is computed from all the data points)
error bands. $\chi^{2}$ is a measure of goodness of fit.
} 
\end{figure}

\subsection{First Rotation Rates : Variations With Respect to Latitude and Number of Observed Days $\tau$ }
In the previous subsections, on the basis of small
magnitude of second coefficient (as illustrated in Figures 3 and 4) in the rotation law 
and the ratio $R$ of CH, we concluded that CH
rotate rigidly. Although second coefficient is small,
it is not completely negligible to conclude unambiguously
that CH rotate rigidly. That means a small contribution to the second coefficient due to differential
rotation can not be ruled out. This result can be interpreted
as follows. The rotation rates of CH presented
in the previous sections are combination of rotation rates
of CH that are anchored at different parts of the interior
during their evolutionary passage on the visible disk.
That means if CH are originated only in the convective envelope
and raised their anchoring feet towards the surface, owing
to differentially rotating convection zone and similar
to magnitudes of rotation rates of sunspots, one
should get a reliable and large magnitude of second
coefficient in the rotation law. On the other hand,
if the CH are originated in the radiative core and
raised their anchoring feet towards surface, during
their first appearance on the surface, one 
should get combined contribution (from the differential and rigidly rotating
regions) to
the second coefficient. That means if one computes
the {\em first rotation rates} $\Omega_{1}$ of CH during their first appearance
on the surface for different latitudes and number of days
 ($\tau$) observed on the disk,  one should get unambiguously negligible contribution
from the second coefficient of the rotation law.
In order to test this conjecture, first rotation
rates $\Omega_{1}$ of CH are computed as follows.
Again, we consider CH that are born between +65$^\circ$ to -65$^\circ$ 
from the central meridian. From the first and second day computed longitudes
(from the central meridian) of CH and by using first method,
{\em first rotation rates} are computed. 
Each {\em first rotation rate} is collected in 5$^\circ$ latitude
bins and average of the first rotation 
rates is computed and, for different latitudes, are illustrated in Fig 9(a).
%Presently we are not understanding why the second
%coefficient of rotation law is positive.
Similarly, for different $\tau$, 
first rotation rates are collected and average of 
first rotation rates is computed and the results are presented in Fig 9(b).
It is important to note that, according to our conjecture, 
we find that magnitude of the second coefficient has 
a negligible contribution to the rotation law that
leads to inevitable conclusion that {\em CH must rotate
rigidly}. 

From all these results, finally we conclude unambiguously that, {\em independent of
their area, number of observed days ($\tau$) and latitude, CH rotate rigidly
during the evolutionary passage on the solar disk}. However, it is
interesting to note from the present and previous studies (Wagner 1975; 
Wagner 1976; Timothy \& Krieger 1975; Bohlin 1977) 
that although whole coronal hole structure rotates rigidly, individual
coronal bright points (CBP) that are embedded in the
coronal holes rotate differentially (Karachik {\em et.al.} 2006). As pointed by
these authors, coronal bright points in the corona might be influenced by the
surrounding differentially rotating plasma. However, it is not clear
 how CBP are influenced by the differential rotation of the surrounding plasma. 

\section{DISCUSSION AND CONCLUSIONS}
In contrast to other persistent solar features
of the corona, then, why do coronal holes rotate rigidly?
Many observations (Madjarska {\em et.al} 2004; Subramanian {\em et.al} 2010;
Tian {\em et.al.} 2011; Yang {\em et.al} 2011; Krista 2011; Krista {\em et.al} 2011; 
Crooker \& Owens 2011; Madjarska {\em et.al} 2012)
suggest magnetic reconnection at the coronal hole boundaries (CHBs) as
the cause of rigid body rotation. 

Pevtsov \& Abramenko (2010) conclude that coronal holes$'$ rotation rate is almost like 
rotation rate of sunspots and the CH are
``analogous to a grass fire, which supports itself by continuously propagating
from one patch of dry grass to the other". 
That means  coronal hole constantly changes its footprint moving 
from one available polarity to the other. This implies that area of coronal hole will 
depends on size of available polarity footprint, and it
can either decrease or increase depending on size of photospheric magnetic field patch.
This also suggests that, on average, difference in the coordinates
at the eastern and western boundaries should remain constant yielding a rigid
body rotation rate (as suggested by the previous studies). 
Thus one can argue that coronal holes are surface phenomena.
If the coronal hole is a surface phenomenon and if it constantly changes its
footprint moving from one available polarity to other, area of coronal hole 
depends on size of available polarity footprint. Hence, area should either decreases
or increases with a result that, on average, area with respect to time must be
nearly constant. In order to test this conjecture, in Fig 10, we illustrate 
measured areas (that are corrected for projection)
of CH that have $\tau$ of 4 days and 5 days (upper panel)
and, 6 days and 10 days (lower panel) respectively.
Dates of occurrence of these individual coronal holes that are presented in
the upper panel of are : 6th Nov to 9th Nov 2001 (2$^{\circ}$ to 41$^{\circ}$ East of central
meridian); 8th May to 12th May 2004 ( 30$^{\circ}$ East to 12$^{\circ}$ west
of central meridian)  and, dates of occurrence of coronal holes that are
presented in the lower panel are: 21st Aug to 26th Aug 2003 (45$^{\circ}$ East to 11$^{\circ}$
west of central meridian); 22nd Dec to 31st Dec 2005 (50$^{\circ}$ East to 62
$^{\circ}$ west of central meridian) respectively.

One can notice from Fig 10 that, contrary to expectation 
(that area of coronal hole nearly remains constant during its evolution),
on average, coronal holes$'$ area smoothly decrease (upper panel) continuously or
increase like sunspots$'$ area evolutionary curve, reaches maximum area and then smoothly decreases (lower panel).
From these figures, we can not find other expected signatures for
the reconnection, {\em viz.,} substantial daily variations of areas of CH
during their evolution. This does not mean that there is no magnetic reconnection at the boundaries.
However, in the following, we show that magnetic reconnection alone
can not be sufficient for explanation of dynamics (rigid body rotation) and
area evolution of the coronal holes. Hence, coronal holes must
be deep rooted rather than mere surface phenomena.
Interestingly, similar to Bohlin's (1977) study,
we also find the same order ($\sim 10^{14} cm^{2}/sec$) of average growth (or decay) of
CH.

%that is almost similar to diffusion rate of super granular cells
%(Leighton 1964) of the convection zone. This result suggests that CH must be deeply rooted, probably 
%below base of the convection zone. 
One would also expect, magnetic reconnection at the boundary of CH might have a substantial
contribution for the enhancement of the average intensity (DN counts). In order
to check this expectation, for the same CH presented in Fig 10,
we compute the daily average DN counts 
($= \frac{\sum\limits_{i=1}^n DN_{i}}{N}$, where N is total number of pixels)
 and are illustrated in Fig 12. Obvious fact from Fig 10 and 12 is that
as the area of CH increases, average DN counts (intensity) decrease
and {\em vice versa}. However, according to our expectation, coronal holes do not 
show any transient and substantial increase in the intensity during their 
daily evolutionary passage on the solar disk.

Off course, as CH is embedded in the
atmosphere where closed field lines due to active regions coexist and hence, it
is natural to expect reconnection at the boundary of a CH
due to oppositely directed field lines. 
Possible reason for the null detection of magnetic
reconnection from our data set is due to low temporal
resolution of daily data used in this analysis. 
In fact, with a high temporal resolution of CH data set, majority of previous
studies (Wang {\em et al.} 1998; Madjarska {\em et al.} 2004; 
Raju {\em et.al} 2005; Aiouaz 2008; Madjarska \& Wiegelmann 2009; 
Subramanian et al. 2010; Edmondson {\em et.al.} 2010; Krista 2011;
Yang {\em et.al.} 2011; Madjarska {\em et.al} 2012) show the evidences of reconnection,
although other studies have lack of such a evidence (Kahler \& Hudson 2002; Kahler et al. 2010).
If we go by the majority of results that rigid rotation rates
of the coronal holes is due to magnetic reconnection at the
coronal hole boundaries, then one would expect that
the shape (area) of the coronal hole during their disk passage must remains
constant. As most of these majority of studies used short ($\sim$ hours)
duration data set, question arises whether CH maintain their shape (and hence
their areas) through out disk passage (as one can see from our analysis, most of CH exist
more than 5 days on the solar disk). One can notice from the area-time plots (Fig 10), 
during ( $\sim$ days) its disk passage,
CH do not maintain their shape and hence rigid body
rotation rate of CH is not due to interchange reconnection.
As the previous studies use high temporal, short duration ($\sim$ hours) data set 
and during such time scales (as the CH has a large dimension) obviously one gets constant shape
and hence conclusion (that rigid rotation rates of CH is due to
magnetic reconnection) is right. However, again we stress from the
results presented in Fig 10 that,
on long duration ($>$ 5 days), CH do not maintain their shape
and rigid body rotation rate of CH is not due to magnetic reconnection
alone at their boundaries. Rigid body rotation rate of CH is likely due to their
deep rooted anchoring of their feet and subsequently raising towards the surface
and then to the atmosphere.

As for area evolution of the coronal hole, question arises as to 
which is the dominant physical process that dictates
temporal variation of area and hence removal of magnetic flux
of the coronal hole? Is it due
to {\em magnetic diffusion} (whose diffusion time
scale is $\sim$ ${L \over \eta^{2}}$) or {\em magnetic reconnection} 
at the coronal hole boundaries?
Similar to sunspots$^{'}$ area evolution curve (Hiremath 2010), formation
and growth part of area evolution of CH are not understood. However,
in order to answer afore mentioned queries, we consider decay
part of the area evolution curve with 
following two physical reasonings:
(i) if area evolution of CH is dominated by magnetic diffusion, then its
area must varies as $\sim$ $t^{-1/2}$ (where $t$ is time variable) and, (ii)
if area evolution of CH is mainly due to magnetic reconnection, annihilation
of magnetic flux due to reconnection of opposite magnetic
field lines at the boundaries of the coronal hole leads to
an exponential decrease of area with time. 
If coronal hole is considered to be
cylindrical magnetic flux tube with uniform magnetic field
structure, from magnetic induction equation (with diffusive dominated term),
it is instructive to show that equation for rate of change of magnetic flux $\phi$ is 
$\frac{d \phi}{d t} = \eta \frac{d^2 \phi}{d z^2}$ (where $\phi$=$\int^r_0 B_{z} A dr$
is magnetic flux of coronal hole flux tube, $A (=2 \pi r^{2})$ is area, $t$ is time variable, $\eta$ is magnetic diffusivity, 
$B_{z}$ is a uniform magnetic field structure along the $z$ direction and $r$ is
radius of flux tube). From the results (Krista and Gallagher 2009; CHARM algorithm
from solarmonitor.org) illustrated in Fig 11a, absolute magnitude
of $B_{z}$ of 10 days CH (during decay part of
its area evolution as presented in Fig 10) is found to be nearly independent of time (number of observed days). Using this observational
information and assumption that magnetic field structure of coronal hole is
also uniform spatially along $r$ direction, it can be easily shown from the rate of change of magnetic flux equation 
that $\frac{d A}{d t} = \eta \frac{d^2 A}{d z^2}$ and
whose solution is obtained as $A \sim t^{-1/2}$ on diffusion time scales.  

In order to test these afore mentioned two reasonings, 
for example, decay part of 10 days area evolution
curve is subjected to diffusion and exponential fits. After 
linearizing the two laws, least-square
fits are performed and the result is illustrated in Fig 11b.
Compared to exponential fit, for the decay part of area evolution curve, least-square fit 
for law of diffusion yields very
low value of $\chi^{2}$ with the expected
decay index of $\sim$ -0.5. 
Hence, during decay part of its evolution of area,
coronal hole is consistent with the first reasoning and area evolution of CH is mainly dictated by
magnetic diffusion.
However, persistent magnetic reconnection at the
boundaries of CH during their evolution can not be neglected.
 Thus, it is reasonable to conclude that 
{\em both the magnetic diffusion and the reconnection processes
control the evolution of area of CH during their passage on the
solar disk}. 

Another important result from this study is that why coronal holes rotate with
a magnitude of $\sim$ 438 nHz during their first appearance, where as other active
regions, approximately at the same height in the corona, have a magnitude
of rotation rate similar to rotation rate of sunspots. Moreover, similar
to sunspots, coronal holes are likely to be three dimensional structures
whose dynamical evolution is not only controlled
by the surface activity, but also related to
the solar interior dynamics where roots
of CH might be anchored, probably below base of convection zone. 
This idea that CH probably might be originated below base
of the convection zone is not a new one. In fact,
nearly three decades back, Gilman (1977) came to the conclusion
that CH$^{'}$ origin and formation may not be due to
so called ``dynamo mechanism" that apparently explains
the genesis of sunspot cycle. 
While discussing the origin of XBP (X-ray bright points),
Golub {\em et. al.} (1981) came to the conclusion that
XBP and coronal holes probably might be originated
below base of the convection zone.
Recently, Jones (2005) also expressed
similar doubt that origin of CH is in the convection zone and concludes
that their roots must be further deeper below
base of convection zone.
Very recently, by investigating the formation of isolated, non-polar
coronal holes on the remnants of four decaying active regions
 at the minimum/early ascending phase of sunspot activity,
Karachik {\em et. al.} (2010) came to a similar conclusion that, during their first
appearance, CH might be deeply rooted. 
%is correct
%in saying that roots of CH are connected below base of the convection
%zone.

Hence, on the basis of these two important results ((i) first rotation rates of
CH during their initial appearance and during evolutionary passage and,
(ii) magnitude of rotation rates ($\sim$ 438 nHz)), we suggest a possibly naive but 
plausible reasonable proposition in the following way.
Compared to other activity indices such as x-ray bright points (XBP),
coronal holes are very large ( $\sim$ 10 times the typical big sunspot) 
and it is not unreasonable to suggest that their roots
may be anchored very deep below the surface. In case of coronal XBP, from the nature of their
differential rotation rates, Hara (2009) has conjectured that their roots might be anchored
in the convective envelope, as helioseismic inferences (Antia et al. 1998; Antia \& Basu 2010) show that whole convective envelope is rotating differentially.
On the other hand, the present and previous studies (Wagner 1975; Wagner 1976; Timothy \& Krieger 1975; 
Bohlin 1977)
strongly suggest that the rotation rate of coronal holes is independent of latitude, 
number of days ($\tau$) observed on the disk and area.
%and rotate rigidly. 

\begin{figure}
%\epsscale{.80}
\begin{center}
\begin{tabular}{cc}
{\includegraphics[width=18pc,height=18pc]{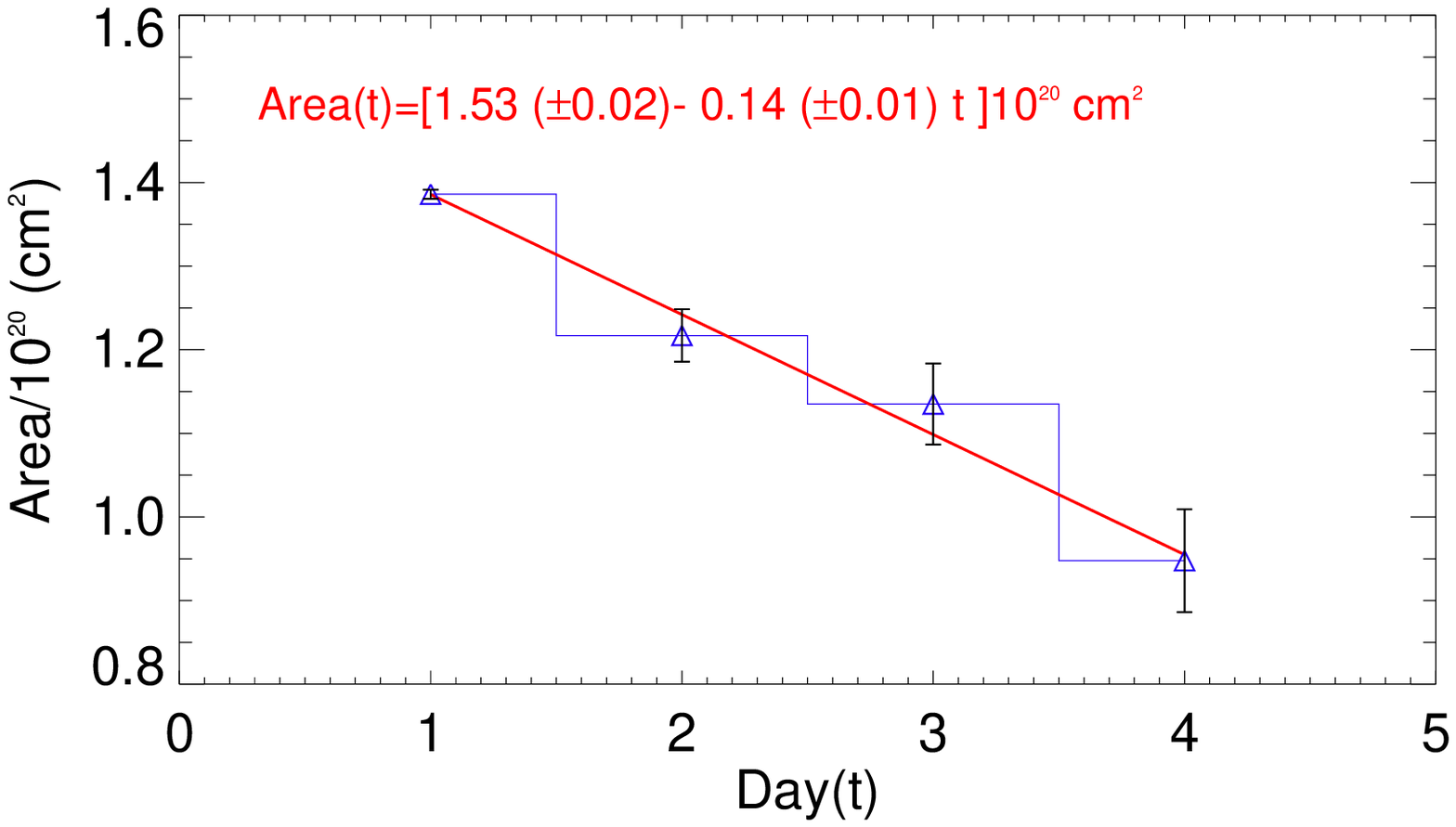}}&
{\includegraphics[width=18pc,height=18pc]{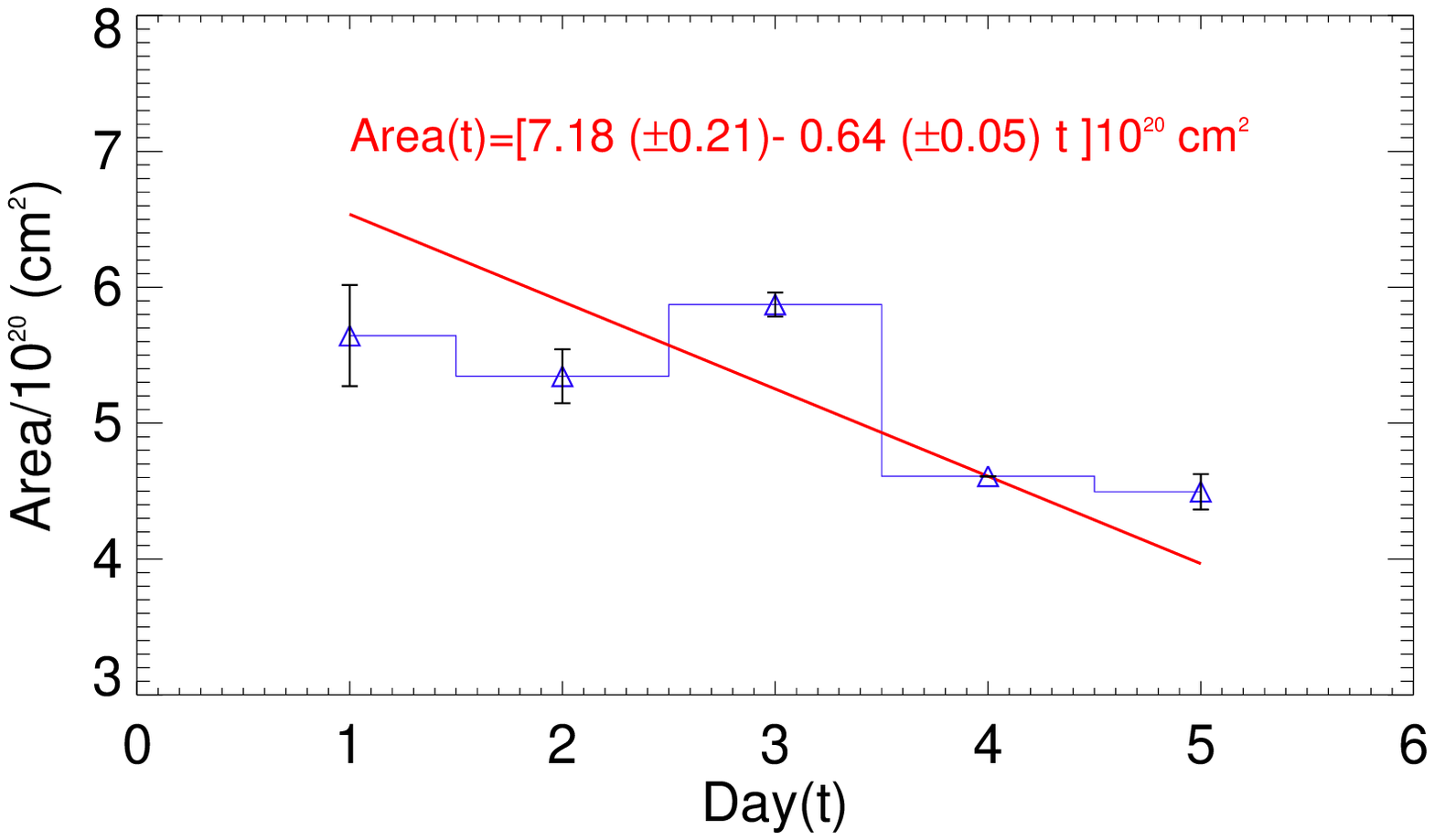}}\\
\end{tabular}
\begin{tabular}{cc}
{\includegraphics[width=18pc,height=18pc]{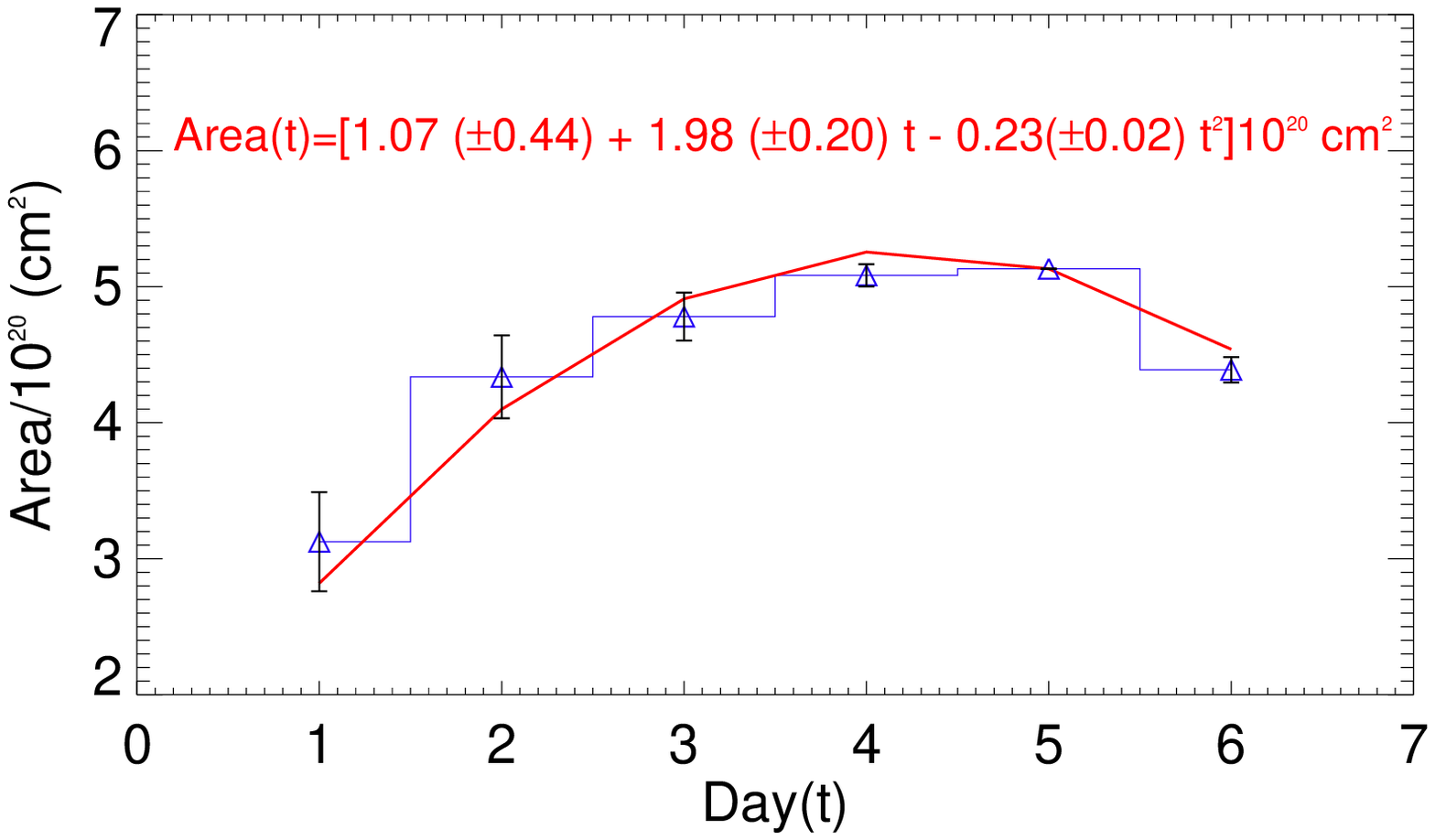}}&
{\includegraphics[width=18pc,height=18pc]{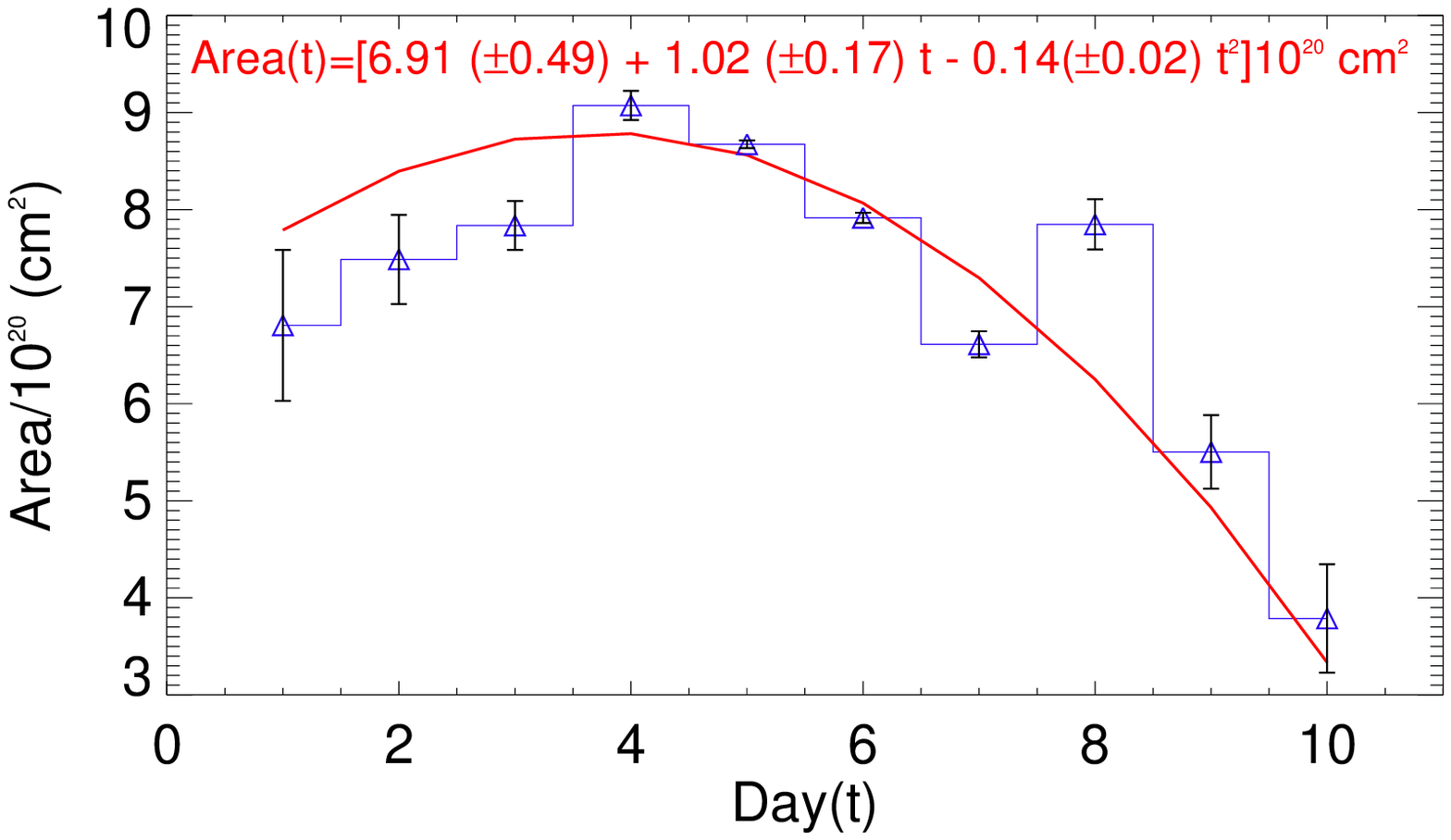}}\\
\end{tabular}
\end{center}
\caption{
For different days, measured average areas of the CH (blue bar plot) that are normalized
with the area $10^{20} cm^{2}$. Figures in the upper panel are the variation
of areas of CH for the number of observed 4 and 5 days ($\tau$) respectively. 
Whereas figures in the lower panel illustrate the variation of areas of CH for 
the number of observed 6 and 10 days ($\tau$) respectively. $\chi^{2}$ is a measure of goodness of fit.
}
\end{figure}

\begin{figure}
%\epsscale{.80}
\begin{center}
\begin{tabular}{cc}
{\includegraphics[width=18pc,height=18pc]{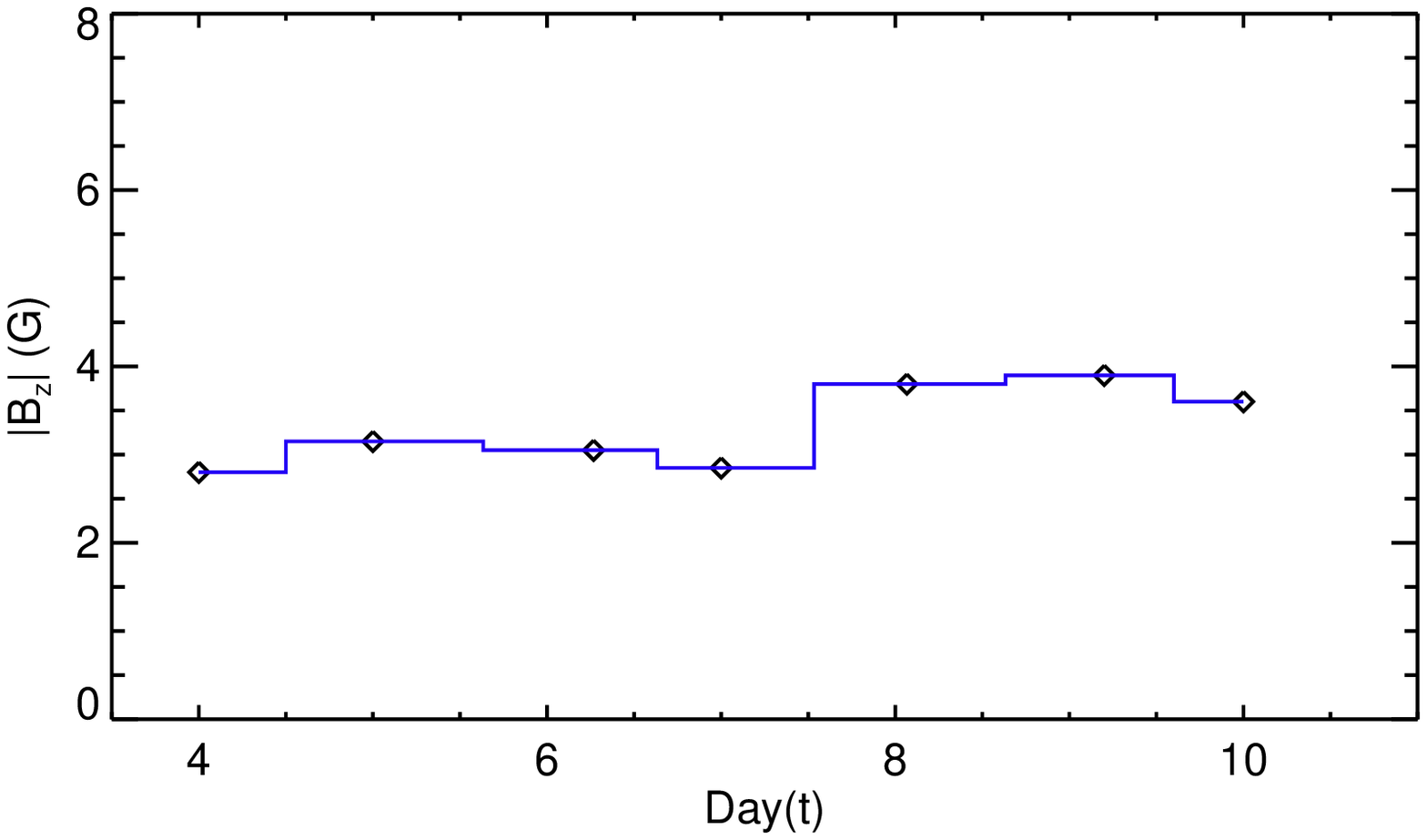}}&
{\includegraphics[width=18pc,height=18pc]{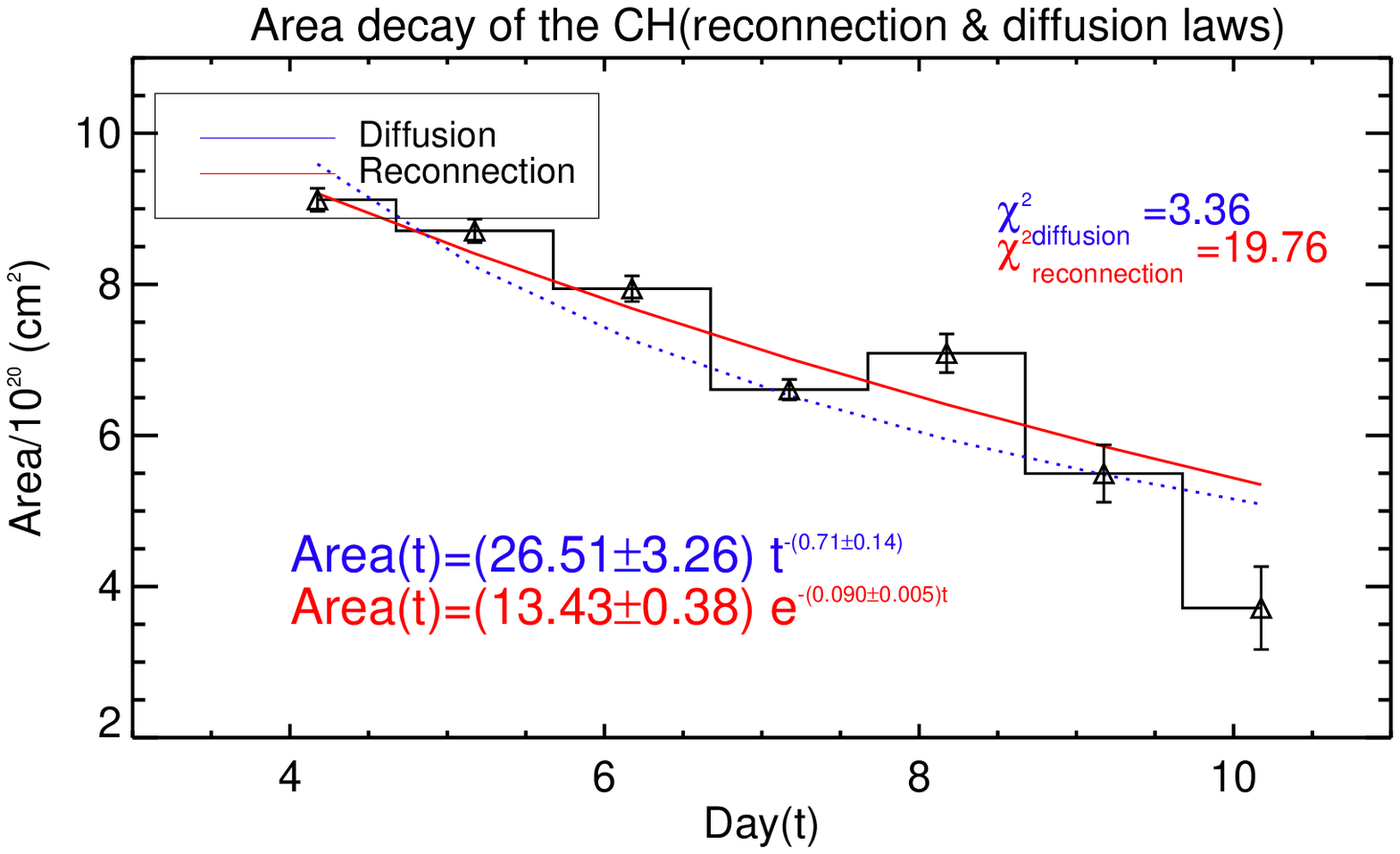}}\\
\end{tabular}
\end{center}
\caption{
Variation of absolute magnitude of magnetic field structure (blue bar plot connected by diamonds; 
left figure) and area (black bar plot
connected by triangles; right figure) of a CH for the decay part of 10 days area evolutionary curve. Areas 
are normalized with the area $10^{20} cm^{2}$ and are subjected
to laws of magnetic diffusion ($Area(t)=A_{0}t^{-n}$, $t$ is time variable) and magnetic reconnection 
($Area(t)=A_{0}e^{-c_{1}t}$) respectively. $A_{0}$, constants $n$ and $c_{1}$
are determined from the least square fit. $\chi^{2}$ is a measure of goodness of fit.
%Small value of $\chi^{2}$ (typically
%$\chi^{2}$ should be $\le$ (N-n), where $N$ is total number of data points
%and $n$ is degrees of freedom, in this case $n=2$) implies fit is very good.
}
\end{figure}

\begin{figure}
%\epsscale{.80}
\begin{center}
\begin{tabular}{cc}
{\includegraphics[width=18pc,height=18pc]{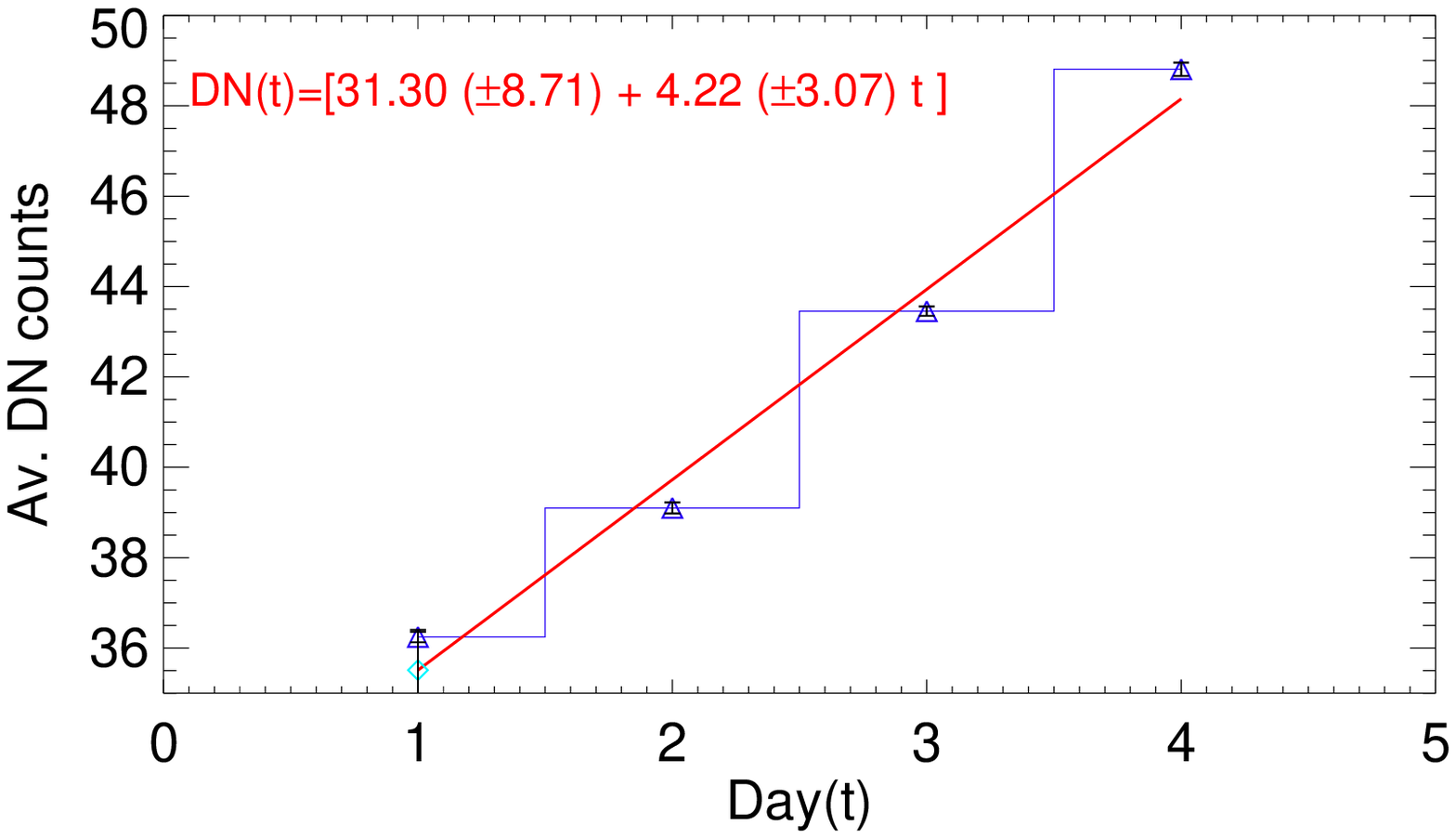}}&
{\includegraphics[width=18pc,height=18pc]{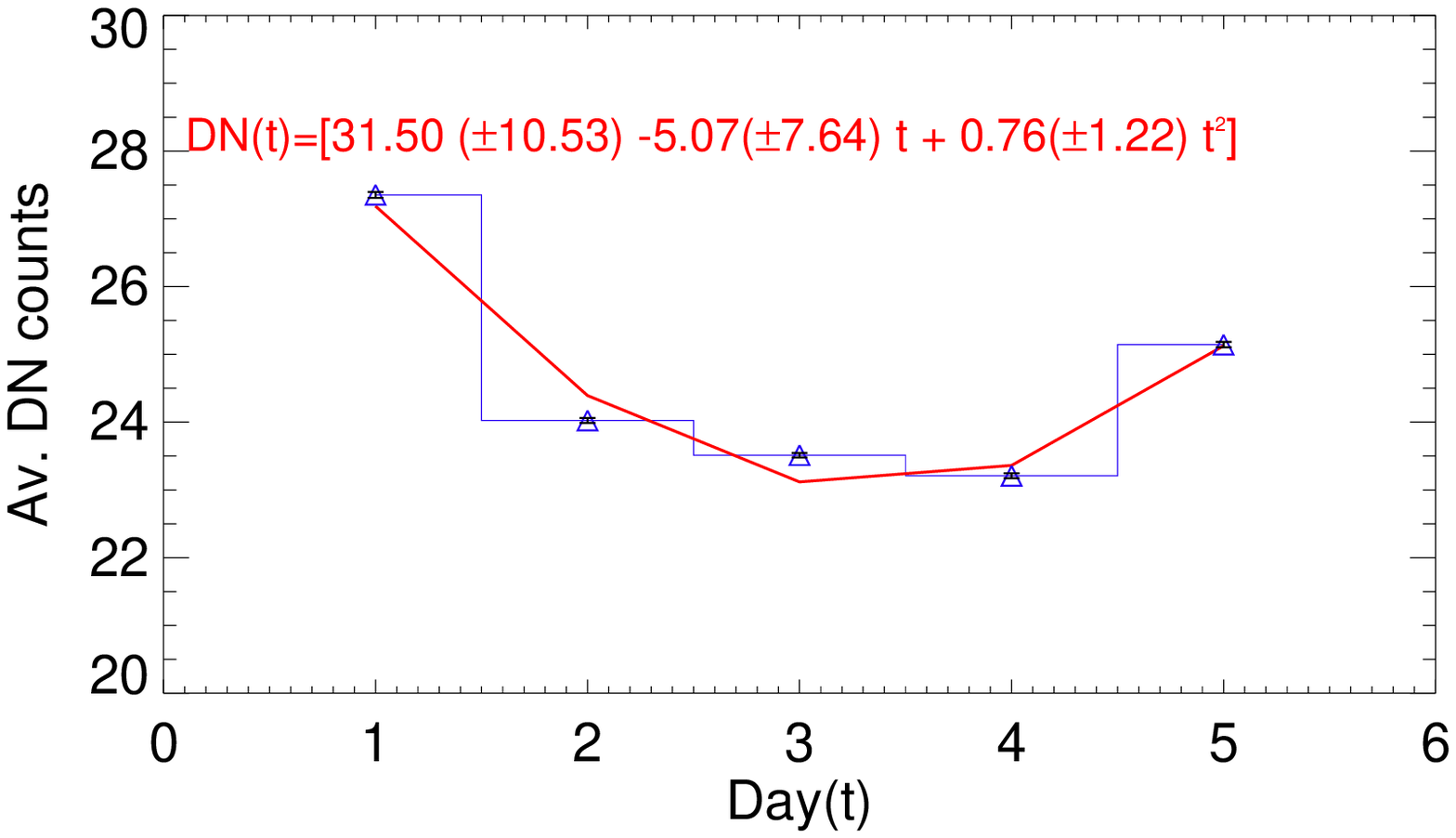}}\\
\end{tabular}
\begin{tabular}{cc}
{\includegraphics[width=18pc,height=18pc]{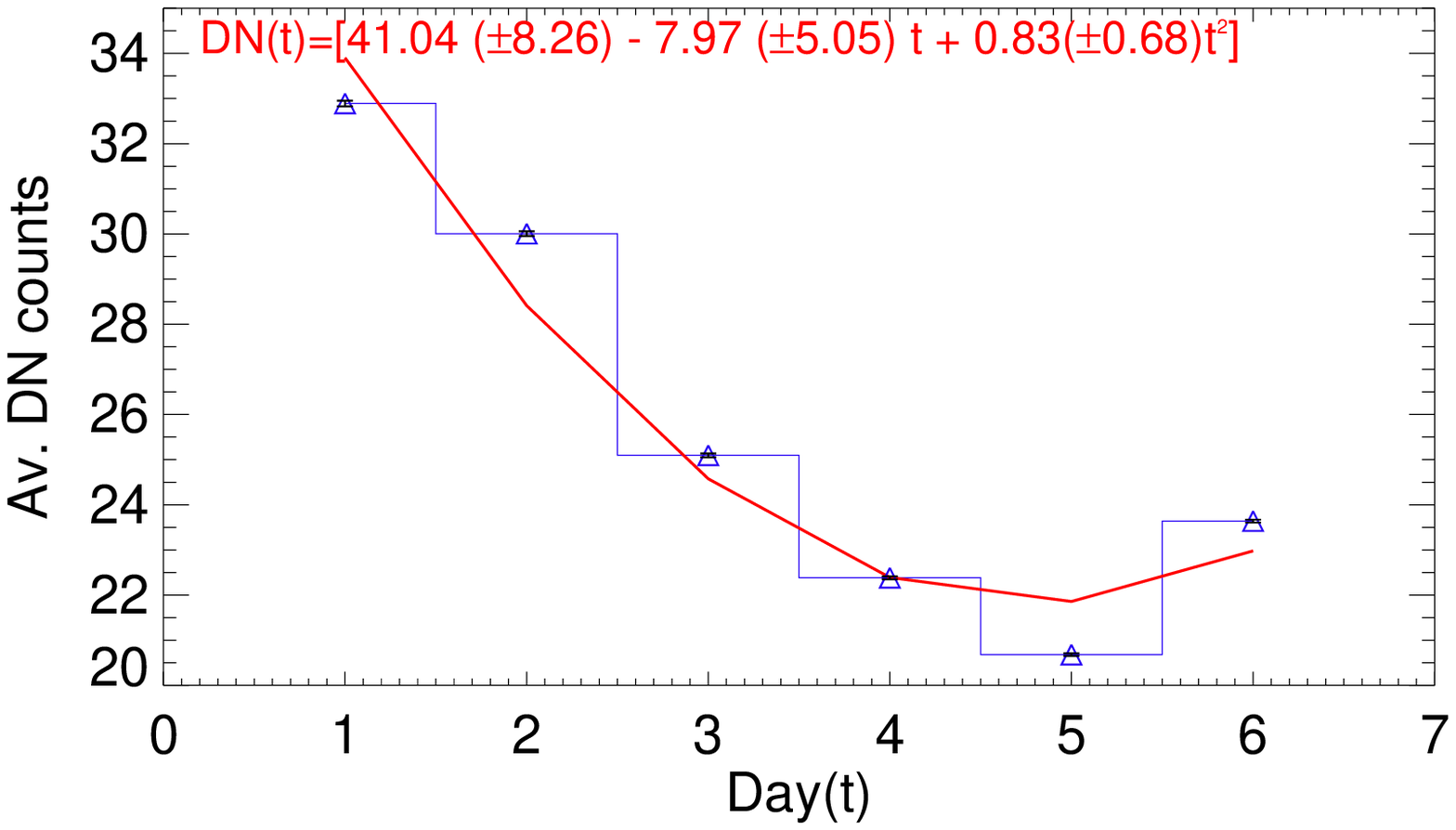}}&
{\includegraphics[width=18pc,height=18pc]{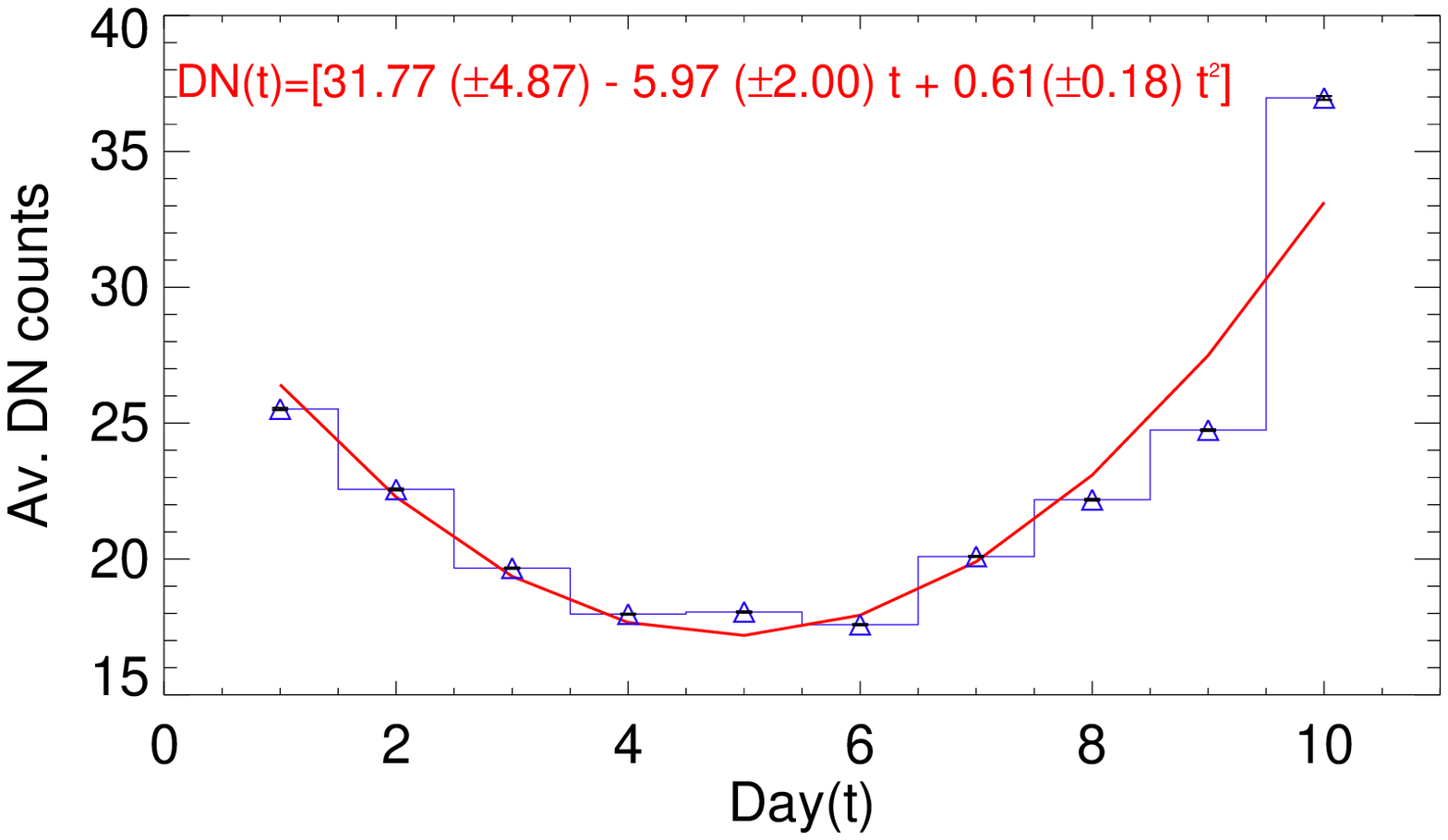}}\\
\end{tabular}
\end{center}
\caption{
For different days, measured average DN counts of the CH (blue bar plot). 
Figures in the upper panel are the variation
of average DN counts of CH for number of observed 4 and 5 days ($\tau$)
 respectively. Whereas the figures in the
lower panel illustrate variation of average DN counts of CH for number of 
observed 6 and 10 days ($\tau$) respectively.
%In all the figures, red continuous line is the polynomial fit of degree 3.
}
\end{figure}

\begin{figure}
%\epsscale{.80}
\begin{center}
\begin{tabular}{cc}
{\includegraphics[width=18pc,height=18pc]{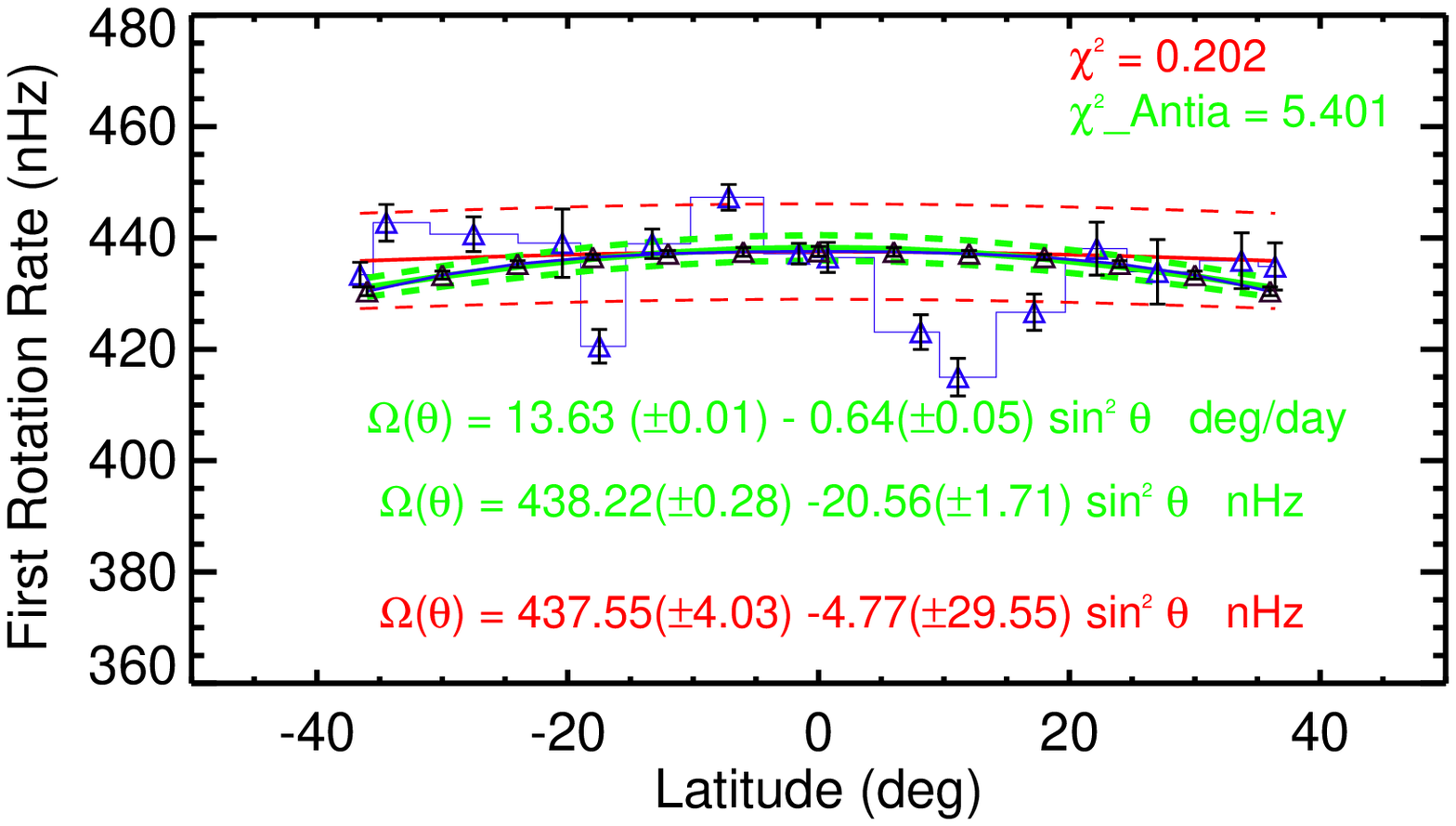}}\\
\end{tabular}
\end{center}
\caption{
For all the sizes and number of observed days ($\tau$), 
figure illustrates first rotation rates of coronal holes (blue bar plot
connected by blue triangles),
 with a least-square fit (red continuous line). Also
plotted is the helioseismically inferred (Antia \& Basu 2010) rotation rate
(green continuous line connected by black triangles with green dashed lines as
one sigma error bands)
at a depth of $0.62(\pm 0.10) R_{\odot}$, as a function of latitude.
Red and green dashed lines represent
one standard deviation (that is computed from all the data points)
error bands. $\chi^{2}$ is a measure of goodness of fit.
}
\end{figure}

As for the anchoring depths, during their first appearance
in the corona and owing to its magnetic nature (Gurman {\em et al.} 1974; Bohlin 1977; 
Levine 1977; Bohlin \& Sheeley 1978; Stenflo 1978; Harvey \& Sheeley 1979; Harvey et al. 1982; 
Shelke \& Pande 1984; Obridko \& Shelting 1989;  Zhang et al. 2006; Fainshtein 2010), 
we expect that a coronal hole might isorotates with the
solar plasma, so its rotation rate during its first appearance
and the rotation rate at the anchoring depth must be identical.
It is interesting to note that the average rotation rate ($\sim$ 438 nHz),
we have measured in coronal holes (Fig 13) is similar to that
of the average rotation rate of the solar plasma inferred by helioseismology
(Antia \& Basu 2010; rotation rate of the solar interior averaged over one solar cycle is kindly provided by Prof. Antia) at
a depth of $\sim 0.62(\pm 0.10) R_{\odot}$. Hence, during
 first appearance of the coronal hole, it is reasonable to suggest that the depth
of anchoring of CH might be around $0.62(\pm 0.10) R_{\odot}$.
If we simply identify the
rotation rates found here with the internal rotation rate at a given depth, we
find a match only inside the radiative interior, at a depth of $0.62(\pm 0.10) R_{\odot}$ 
solar radii. In future, helioseismology may give further inferences
on the anchoring depths of coronal holes. We know, however, of no currently accepted model of magnetic field
generation that could anchor coronal structures to such a depth in the
interior. With a caveat that unless a consistent
and acceptable theoretical model of CH that supports of our proposition
(that during their first appearance, roots of CH might be anchored
in the radiative core), our proposed idea remains mere a conjecture only.\\

To conclude this study, we used SOHO/EIT 195 $\AA$  calibrated  images to
determine the latitudinal and day to day variations of rotation rates of the
coronal holes. We found that: (1) irrespective
of their areas and number of days ($\tau$) observed on the disk, for different latitude
zones, rotation rates of  CH follow a rigid body rotation law,
(2) CH also rotate rigidly during their evolution history and,
(3) during their first appearance, CH rotate  rigidly with a constant angular
velocity $\sim$ 438 nHz which only matches depth around $0.62(\pm 0.10) R_{\odot}$, in the radiative interior.
This result is so counterintuitive that we can only conclude that we do not understand why CH rotate rigidly at that rate.

%\acknowledgments
\centerline{\bf Acknowledgements}

Authors are grateful to an anonymous referee for the invaluable 
comments and suggestions that substantially improved the
results and presentation of the manuscript.  
Authors are also grateful to Dr. J. B. Gurman for giving useful information
on the SOHO data, for going through the earlier version of
this manuscript and, for giving useful ideas.
Hiremath is thankful to former Director, Prof. Siraj Hasan,
Indian Institute of Astrophysics, for encouraging this ISRO funded project.
This work has been carried out under  ``CAWSES India Phase-II
program of Theme 1'' sponsored by Indian Space Research Organization(ISRO),
Government of India. SOHO is a mission of international cooperation between ESA and NASA.

\end{document}